\documentclass[epj]{svjour}
\usepackage{graphics}

\usepackage{amssymb}
\usepackage{amsmath}
\newcommand{\Efield}{\mathcal{E}}
\newcommand{\au}{\mathrm{a.u.}}

\newcommand{\ver}{\mathbf{r}}

\newcommand{\vk}{\mathbf{k}}

\usepackage{xcolor}
\begin{document}
\title{Quantum interference and imaging using intense laser fields.}
\author{Kasra Amini \inst{1,2} \and Alexis Chac\'{o}n \inst{3,4} \and Sebastian Eckart\inst{5} \and  Benjamin Feti\'{c} \inst{6} \and Matthias
K\"{u}bel \inst{7}
\thanks{matthias.kuebel@uni-jena.de}%
}                     
\offprints{}          
\institute{
ICFO -- Institut de Ciencies Fotoniques, The Barcelona Institute of Science and Technology, 08860 Castelldefels (Barcelona), Spain
\and
Max Born Institute, 12489 Berlin, Germany
\and
Department of Physics and Center for Attosecond Science and Technology, POSTECH, 7 Pohang 37673, South Korea
\and
Max  Planck  POSTECH/KOREA  Research  Initiative,  Pohang  37673,  South  Korea
\and
Institut f\"{u}r Kernphysik, Goethe-Universit\"{a}t, Max-von-Laue-Str. 1, 60438 Frankfurt am Main, Germany
\and
Faculty of Science, University of Sarajevo, Zmaja od Bosne 35, 71000 Sarajevo, Bosnia and Herzegovina
\and
Institute of Optics and Quantum Electronics, Max-Wien-Platz 1, D-07743 Jena, Germany
}
\date{Received: date / Revised version: date}
%
\abstract{
The interference of matter waves is one of the intriguing features of quantum mechanics that
has impressed researchers and laymen since it was first suggested almost a century ago. Nowadays, attosecond
science tools allow us to utilize it in order to extract valuable information from electron wave packets. Intense laser
fields are routinely employed to create electron wave packets and control their motion with attosecond and \aa{}ngstr\"{o}m precision. In
this perspective article, which is based on our debate at the Quantum Battles in Attoscience virtual workshop 2020, we discuss some of the peculiarities of intense light-matter interaction. We review some of the most
important techniques used in attosecond imaging, namely photoelectron holography and laser-induced electron diffraction. We
attempt to ask and answer a few questions that do not get asked very often. For example, if we are interested in position space
information, why are measurements carried out in momentum space? How to accurately retrieve photoelectron spectra from the numerical
solution of the time-dependent Schr\"{o}dinger equation? And, what causes the different coherence properties of high-harmonic
generation and above-threshold ionization?
} 
\maketitle

\section{Introduction}
\label{intro}
\sloppy

Scientific progress has been fueled by the dream to visualize objects or phenomena that are either too small or too
fast to be directly perceived by our senses. For example, femtosecond laser pulses offer the opportunity to freeze
femtosecond dynamics. This capability has provided unique insights into ultrafast processes such as chemical
reactions \cite{Zewail2000}. Moreover, attosecond electron dynamics have been resolved with the help of short-wavelength
attosecond light sources \cite{Corkum2007,Krausz2009,Vrakking2014}. Alternatively, the time resolution can be pushed beyond
the femtosecond duration of a laser pulse by using interferometric techniques, e.g.~\cite{Iaconis1998,Lindner2005,Shafir2012}.

\begin{figure}[t]
    \centering
    \resizebox{0.5\textwidth}{!}{
    \includegraphics{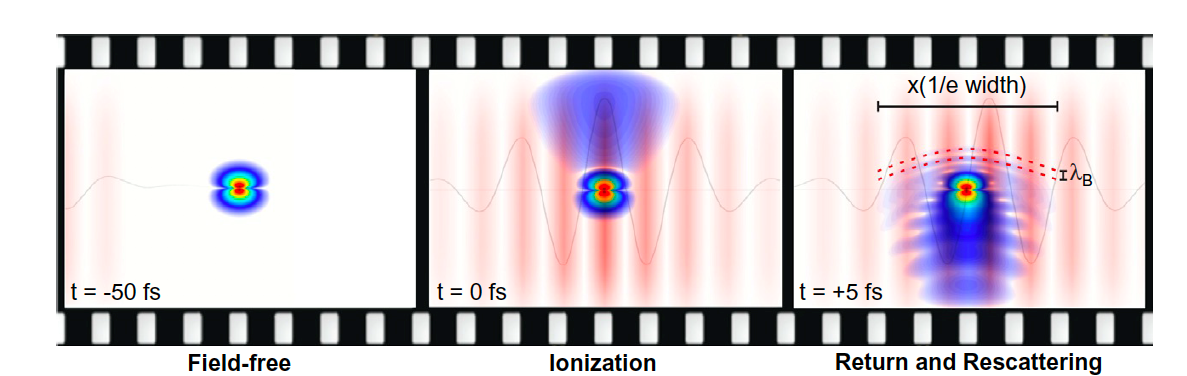}}
    \caption{Quantum mechanical illustration of the creation and rescattering of a photoelectron wave packet in an intense
    infrared laser field. Calculated using QProp \cite{Bauer2006}. Figure and caption reproduced from Ref.~\cite{Amini2020}. }
    \label{fig1}
\end{figure}

This article is a follow up to Quantum Battles in Attoscience 2020 virtual workshop and focuses on Quantum Interference and Imaging
of molecular structure by means of photoelectrons. The pivotal idea is to image atomic-scale structures by utilizing
electrons to overcome the $\sim 1\,\mu$m diffraction limit of infrared laser light. This is achieved by exploiting the
high intensity of the laser pulses to create coherent electron wave packets (EWPs) that are driven by the laser field. With
their short de Broglie wavelength, electrons allow one to push the spatial resolution to \aa{}ngstr\"{o}m scales, while
maintaining the femtosecond (or even attosecond) time resolution dictated by the highly non-linear light-matter interaction. This concept is
depicted in Fig.~\ref{fig1}. Here, the laser-created electron wave packet diffracts upon recollision \cite{Corkum1993} with the parent ion,
encoding information about the scattering potential.

The unique combination of ultrahigh spatial and temporal resolution makes intense laser pulses extremely attractive for
time-resolved imaging. In the present article, we shall focus on techniques that rely on the direct detection of the photoelectrons. The
capabilities of this approach can be extended by exploiting high-harmonic generation (HHG), which allows one to transfer the favorable
properties of the laser-driven electron wave packet into a beam of high-energy photons. The high-harmonic beam can be directly
analyzed or utilized in secondary experiments; both approaches have been extremely fruitful but are beyond the scope of the present paper.  

The underlying process for the creation of electron wave packets using infrared light is strong-field ionization. Its
hallmark feature is the appearance of a series of so-called above-threshold ionization (ATI) peaks in the photoelectron energy spectrum,
spaced by the photon energy \cite{Agostini1979}. An intuitive explanation for these peaks is given in the photon picture: in
order to overcome the ionization potential $I_P$, an atom may need to absorb $n$ photons. Because of energy conservation
this leads to a discrete photoelectron energy $\hbar\omega$, $E_0=n \hbar \omega-I_P$. However, if the field contains
sufficiently many photons, the atom may also absorb $m$ ($m = 0, 1, 2, 3, 4,\mathrm{...}$ ) excess photons,
leading to a series of discrete energies, $E_m=(n+m)\hbar \omega - I_P$. The companion process of HHG can be interpreted in an analogous manner. Here, selection rules require the number of absorbed photons to be odd, such that a comb of only odd photon energies at $E_m=m\hbar \omega$ $(m = 1,3,5, ...)$ is observed.

The ATI peaks can also be understood in the wave picture as a result of quantum interference: at each field
maximum an EWP is created and driven by the laser field. All wave packets will eventually interfere
on the photoelectron detector. Because of the field periodicity $T=\frac{2\pi}{\omega}$, and because time and energy are
conjugate quantities, this interference can be observed in the photoelectron energy spectrum as a modulation with
periodicity $\hbar \omega$. Incidentally, the HHG peaks can be interpreted to be a result of quantum interference, as well. In this case, the EWP is driven back to parent ion and recombination leads to the emission of a photon burst \cite{Corkum1993}. Since equivalent recollisions occur twice per laser cycle, interference of the photon bursts spaced in time by $T/2$, leads to a modulation of $2 \hbar \omega$ in the high harmonic energy spectrum.
Close inspection of the electron dynamics in an intense laser field shows that there
exist, in fact, two instances in each laser cycle which yield the same electron drivt momentum. This leads to another
interference feature, so-called intra-cycle interference whose periodicity varies throughout the photoelectron
spectrum \cite{Bergues2007,Arbo2010}. Importantly, the periodicity of the intra-cycle interference fringes is
always larger than $\hbar \omega$, because the difference of the responsible ionization times is always smaller than $T$. Hence, the
intra-cycle interferences create a superstructure on the ATI comb in the energy domain. This raises the questions: if the ATI comb
corresponds to photons, what is the corresponding quantity of intra-cycle interferences? Do photons exist on sub-cycle time-scales?

In the past two decades, ATI experiments have progressed from 1D energy-domain to 3D momentum-space measurements.
This has been a fruitful path since ATI is rich of interference features in the spatial domain, as well. Analogously to the
arguments above, these spatial interference features will manifest in the Fourier domain, i.e., in momentum space. Before we
examine such features in detail, we shall discuss in chapter \ref{mom_and_not_pos} why measurements are, in fact, conducted
in momentum space rather than position space; and, equally importantly, review some techniques used to carry out momentum
space measurements, and results obtained therewith. 


While measurements are performed in momentum space, our position-space minds desire position space results.  In order to
retrieve a position space image from a momentum space measurement via Fourier transform one requires the phase of the
momentum wave function, which cannot be directly measured. In chapter~\ref{sec:wavepacketphase}, we address the question
of how phases can be measured in the lab through quantum interference, and discuss an example for reconstructing bound
wave functions by holographic interference. 
By interfering an unknown signal wave with a (known) reference wave, a hologram is created. The concept of holography has
been applied to strong-field photoelectron spectroscopy: electron trajectories that scatter from the nucleus (signal) may
interfere with trajectories that do not scatter (reference). The resulting hologram, i.e., the interference pattern in the
photoelectron momentum distribution, may encode information on the scattering potential at the time of rescattering
\cite{Lein2002,Yurchenko2004,Spanner2004,Huismans2011,Meckel2014,Walt2017,Porat2018}. 

The process of rescattering itself alters quantum interference and encodes structural information of the target. In
chapter \ref{sec:LIED}, we discuss laser-induced electron diffraction (LIED), where an electron wave packet scatters from a molecule to
create a diffraction pattern from it. The resulting diffraction pattern can be described as a superposition of the signal resulting
from several point-scatterers at the internuclear distance $R$. If the electron wavelength is sufficiently short, the internuclear
distance may be retrieved from the diffraction pattern \cite{Meckel2008}. Moreover, exploiting the intrinsic
delay between ionization and rescattering, LIED can be seen as a pump-probe experiment, which has been used to
probe nuclear motion not only in diatomic \cite{Blaga2012} but also polyatomic molecules \cite{Wolter2016,Amini2019}. Finally, electron
diffraction without rescattering can probe electronic structure \cite{Meckel2008} and dynamics \cite{kubel2019spatiotemporal}.

For the meaningful interpretation of experiments, it is often essential that experiment and theory go hand in hand. The
gold standard in the field of quantum dynamics is the time-dependent Schr\"odinger equation (TDSE), ideally in all three
dimensions \cite{Armstrong2021}. Various implementations of the TDSE have been realized, specifically for the problem of
intense light-matter interactions, see, e.g. Ref.~\cite{Becker2002}. However,
 a time propagation
$\displaystyle \vert\psi(t_f)\rangle = U(t_i,t_f)\vert\psi(t_i)\rangle$
of the multi-dimensional wave function is only half the battle. The other
half is retrieving the physical observable of interest from the final wave function. Typically, that is the
unbound part of the modulus-squared of the momentum-space wave function,
i.e. $\vert \phi_\mathrm{free}(t_f)\vert ^2$,
representing the
photoelectron momentum spectrum.  In chapter \ref{sec:benjamin}, we illuminate this particular problem that is imperative for the comparison
of experimental and numerical results.

In chapter \ref{sec:decoherence}, we shall discuss the fundamental limitation of all ultrafast imaging methods, namely,
decoherence. This occurs, for example in complex physical system where coupling to the environment can lead to the loss of
coherence. It is the unspoken necessity of any attempt to resolve quantum dynamics that the dynamics are coherent with the
exciting laser field. It is insightful to examine the coherence properties of laser-driven processes. For example, in ATI,
the electron wave packets emitted by different atoms do not interfere with each other; i.e., interference takes place on the
single-atom level. In HHG, on the other hand, all atoms in the focal volume radiate coherently. What is
the underlying reason for this fundamental difference of these closely related effects? 

The final chapter is dedicated to recent efforts to expand strong-field physics and related imaging techniques to
the condensed phase, particularly quantum materials. These systems can exhibit pronounced coherence effects, and
decoherence plays an important role. One key feature of solids, as compared to gases, is the periodicity of the
binding potential. This has far-reaching consequences, leading to new quantum mechanical effects to be investigated with the ultrafast imaging toolbox.

Atomic units (a.u.; $\hbar=1$, $4\pi\varepsilon_0=1$, $e=1$, and $m_e=1$) are used throughout the paper, unless otherwise stated.

\section{Position and Momentum space}
\label{mom_and_not_pos}
\subsection{Why measure momentum and not position?}

The position of a bound electron in an atom (e.g. atomic hydrogen) is known to be in
the vicinity of its ionic core. Thus, the average momentum of the electron relative to the ionic
core must be zero because otherwise the electron would move away from the ionic core and the electron could
not be bound. Therefore, the expectation value of the momentum is zero, $\left\langle {\bf p}\right\rangle = 0$. However, the
electron possesses non-vanishing kinetic energy, i.e., $\left\langle {\bf p}^2\right\rangle > 0$.

Upon ionization (e.g. because the atom is irradiated with an intense laser pulse) the electron is ejected from the atom and
the liberated electron's position coordinate relative to its parent ion changes as a function of time. The liberated electron
can be modeled by a wave packet that evolves with time. Since the electron wave packet carries valuable information of the
physical system and its dynamics, its characterization is at the very heart of many approaches to study light-matter interaction. This
gives rise to an important question: How to characterize the wave function of a freely propagating electron?

Before we answer this question it is important to be aware that position and momentum are conjugate variables and that the complex valued
wave function's in position space and momentum space are linked by Fourier transformation. This implies that a given electronic state can
be fully expressed by using only position space or by using only momentum space coordinates. Despite this equivalence of position and
momentum space, there is a fundamental difference comparing momentum and position space when it comes to a freely propagating electron:
the momentum of a freely propagating electron is conserved but the position of this electron changes as a function of time. Although this
appears to be trivial from a theoretical perspective, it has far reaching consequences regarding the measurement
of freely propagating electrons in real experiments.

\begin{figure}[t!]
 \centering
 \resizebox{0.45\textwidth}{!}{
  \includegraphics{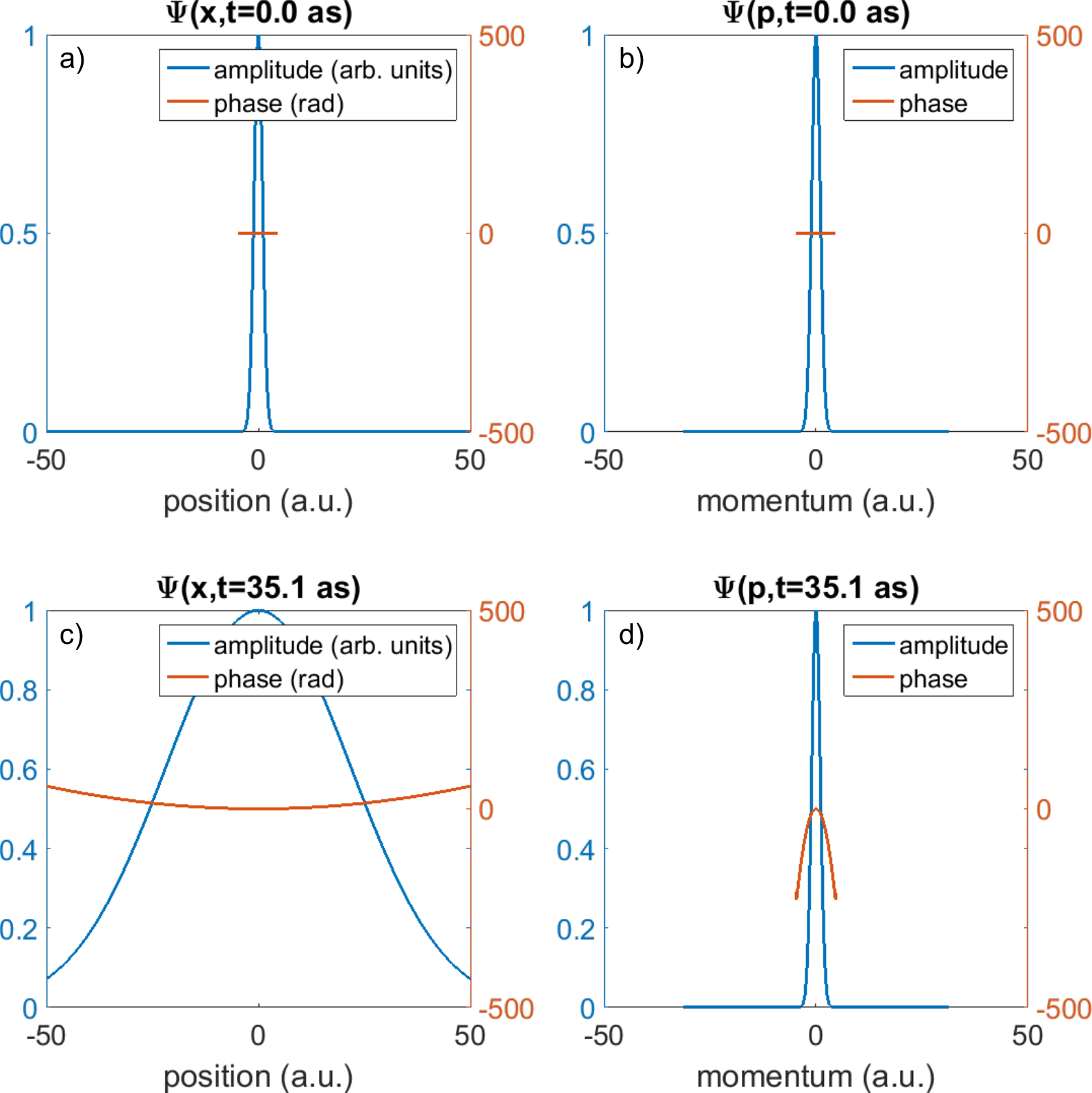}}
  \caption{(a) shows the wave function $\Psi({\bf x},t)$ of a Fourier-limited electron wave packet
  in position space for a given time $t=0$\,as. (b) shows the same as (a) in momentum space. Letting the
  wave packet evolve for about $35$\,attoseconds and without applying any forces to the wave packet leads to (c) and (d), which show the
  wave function in analogy to (a) and (b). In positions space, dispersion leads to a broadening of the distribution of the amplitudes
  whereas the amplitudes in momentum space do not change. Idea taken from a
  manuscript by H. Schmidt-Böcking, H. J. Lüdde, G. Gruber, S. Eckart and T. Jahnke}
  \label{fig:1seb}
 \end{figure}

Let $\Phi({\bf p},t)$ be the complex valued electron wave function in momentum space that depends on the time $t$ and the
three-dimensional momentum ${\bf p}$. In full analogy $\Psi({\bf x},t)$ is the complex valued electron wave function in position space.
We exemplify the relationship  of position and momentum space wavefunctions for a free electron in Fig. \ref{fig:1seb}. $|\Psi({\bf x},t)|^2$ is
time-dependent and evolves on ultrafast time-scales whereas $|\Phi({\bf p},t)|^2$ is time-independent. This directly shows that the position
space distribution has to be characterized as a function of time. In contrast, $|\Phi({\bf p},t)|^2$ is constant. Thus, for
a freely propagating wave packet, the expression $|\Phi({\bf p})|^2$ is useful without specifying at which time it has been measured.

\subsection{Example: Measuring 3D momentum distributions}
\label{section_coltrims}

\begin{figure}[t!]
 \centering
 \resizebox{0.45\textwidth}{!}{
  \includegraphics{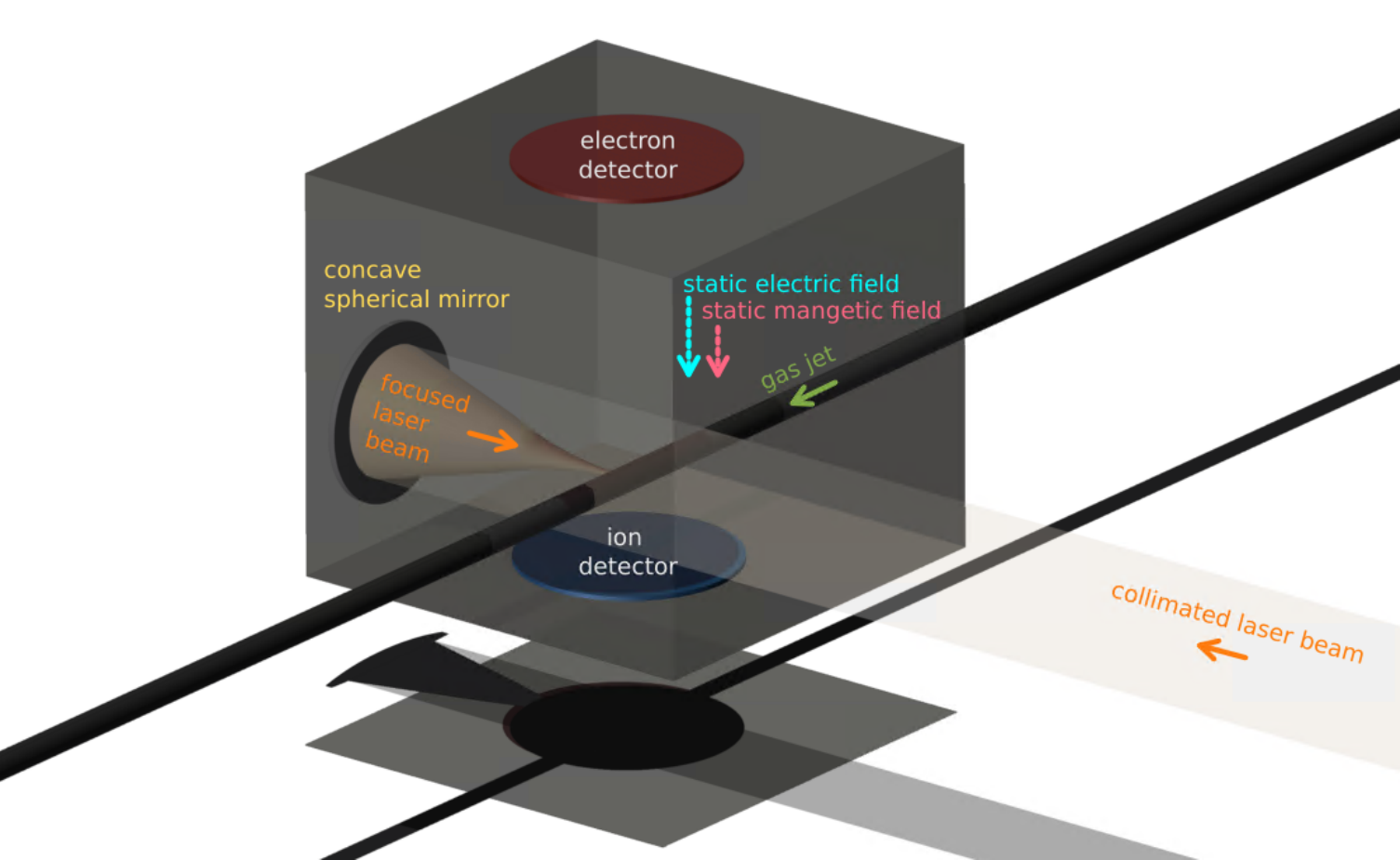}}
  \caption{The basic idea of a COLTRIMS reaction microscope is illustrated. The light-matter
  interaction occurs in the volume in which the focused laser beam and the gas jet overlap.
  The homogeneous electric and magnetic fields are used to guide charged particles, which are
  created in the interaction volume, towards position- and time-sensitive detectors. The gray box
  highlights the volume in which the charged particles propagate on their way from the interaction region
  towards the electron or the ion detector. Figure and caption are taken from \cite{EckartPhDThesis} and
  have been modified.}
  \label{fig:2seb}
 \end{figure}

Despite the theoretical considerations, how to measure the absolute square of the wave
function in momentum space $|\Phi({\bf p})|^2$ in real experiments? The state-of-the art method is to use a
COLd Target Recoil Ion Momentum Spectroscopy (COLTRIMS) reaction microscope \cite{jagutzki2002multiple,ullrich2003recoil} which
makes use of the dispersion of the wave packet in position space that is illustrated in Fig. \ref{fig:1seb}. Allowing the wave
packets to evolve with time (typically for several nanoseconds instead of several attoseconds as in Fig. \ref{fig:1seb}) in the
presence of static external electric and magnetic fields results in a macroscopic distribution of the electron wave packet
that has a size of several millimeters when the wave packet hits a time- and position-sensitive detector (see Fig. \ref{fig:2seb} for
an illustration of a COLTRIMS reaction microscope). The position and the time are typically measured with a precision of several
tens of micrometers and several 100 picoseconds. The additional knowledge about the initial time (typically with a precision
on the order of 100 picoseconds) and position (typically with micrometer precision) of the electron in the spectrometer
allows for the reconstruction of the three-dimensional momentum distribution  $|\Phi({\bf p})|^2$ with a resolution of
typically 1/100 atomic units.

\begin{figure}[t!]
 \centering
 \resizebox{0.45\textwidth}{!}{
  \includegraphics{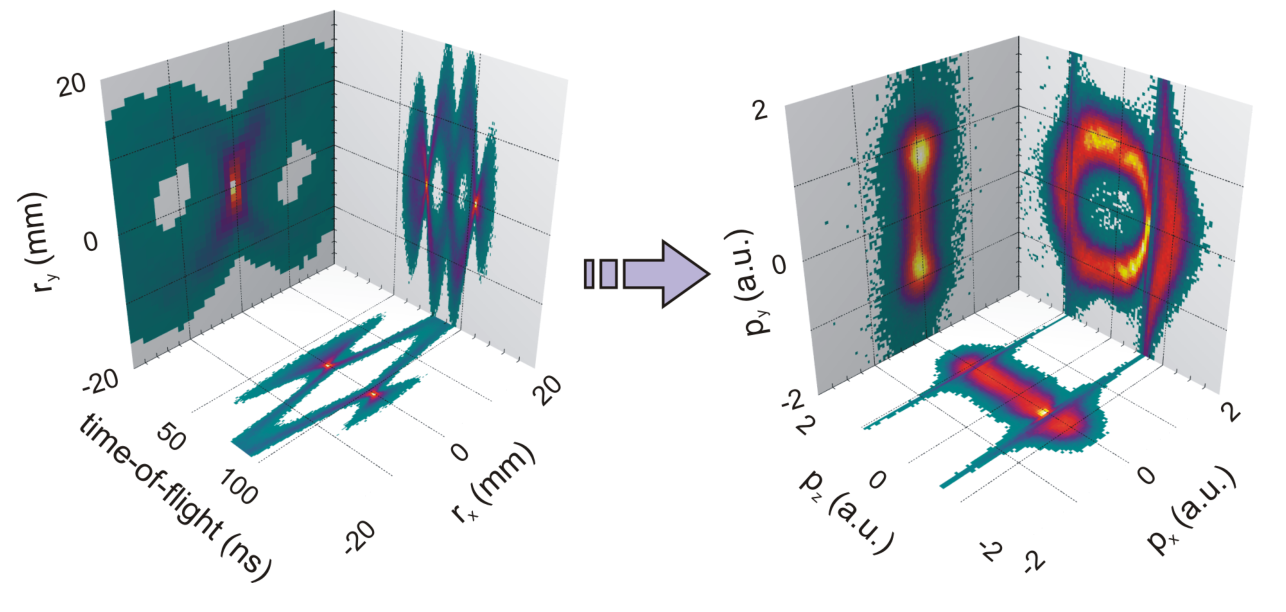}}
  \caption{The 3D detector information (left) can be used to calculate the 3D momentum
  distribution (right) if the geometry of the COLTRIMS spectrometer and the external electric
  and magnetic field are known. The mapping is done by solving the classical equations of
  motion (making use of the knowledge of the focus position in the chamber, the arrival time of
  the laser pulse, and the time and position when the detected particle hits the detector). Figure and
  caption are taken from \cite{AndreStaudte} and the caption has been modified.}
  \label{fig:3seb}
 \end{figure}

This mapping of macroscopic position and time information (nanoseconds and millimeters) to momenta
on the atomic scale is illustrated by Fig. \ref{fig:3seb}. The same conceptual idea underlies the
widely used technique of velocity map imaging (VMI) \cite{eppink1997velocity}. However, while VMI
is similar to COLTRIMS, it (usually) does not resolve the time-of-flight of the particles, resulting in
2D projections of 3D momentum space. However, it should be noted that, in fact, both techniques measure
velocities and not momenta. Moreover, both techniques can be applied to measure not only electrons but also ions.

In summary, for the momentum spectroscopy of liberated electrons it is made use of the
fact that the amplitudes in momentum space do not change as a function of time because
momentum is conserved for a freely propagating particle.

\subsection{Amplitude information in electron momentum space}
\label{amplitude}

What can be learned from the measured electron momentum distribution $|\Phi({\bf p})|^2$?
One famous example is the idea of the
attoclock \cite{Eckle2008,torlina2015interpreting,Teeny2016,sainadh2019attosecond}: Using an elliptically
polarized single-cycle laser pulse it is assumed that the most probable time at which the electron starts
to tunnel \cite{Keldysh1965} is at the maximum of the laser electric field. Further, the final electron
momentum is considered to equal the integral of all laser-induced forces acting upon the electron
after tunneling. Then, the final electron momentum of the electron can be used to retrieve the
time at which the electron appeared at the exit of the tunnel. By evaluating the rotation of the
final momentum distribution with respect to the polarization ellipse, the attoclock has been
used to investigate the time the electron spends inside the tunnel. This interpretation has
led to an ongoing debate \cite{Hofmann2021}, also because of conceptional difficulties regarding the bound
part of the electron wave function \cite{Kunlong2018}, non-adiabaticity \cite{Ni2018_theo} and
the long-range Coulomb-interaction of the electron and its parent ion \cite{Bray2018}. Recent
experiments have set an upper limit of 1.8 attoseconds to the tunneling delay time that is measured
by the attoclock upon the strong field ionization of atomic hydrogen \cite{sainadh2019attosecond}. Further examples
that study light-matter interaction by interpreting amplitudes in final momentum space are
found in Refs. \cite{Alnaser2004,EckartNatPhys2018,Eckart2018_Offsets,Kunlong2019,kubel2019spatiotemporal}.


\subsection{Momentum space imaging of electronic orbitals}

Even in the absence of rescattering, the photoelectron momentum distribution (PMD)
may encode structural information of its origin. The measured far-field photoelectron
momentum distribution  can be understood as a diffraction image of the source. Thus, in principle, it should be
possible to retrieve structural information by analyzing the diffraction pattern. However, the source
is not identical to the atomic or molecular orbital from which the electron is removed but rather the
corresponding Dyson orbital. In addition, the details of strong-field ionization have a decisive impact. The momentum
distribution along the laser polarization is, of course, determined by the time-dependent laser field, leaving the
perpendicular momentum distribution for potential imaging applications. These are, however, hampered by a
markedly distorting filtering effect of tunnel ionization, which strongly suppresses large momenta in the
direction perpendicular to the laser polarization. Specifically, the perpendicular-momentum wave function
at the tunnel exit ($z = z_\mathrm{ex}$) is related to the one at the tunnel entrance, i.e. the Dyson
orbital ($z = z_\mathrm{in}$) by \cite{Murray2010,Murray2011}
\begin{equation}
    \Phi (p_x, p_y,z_\mathrm{ex}) \propto \Phi (p_x, p_y,z_\mathrm{in}) \cdot \exp{\left[-\frac{(p_x^2+p_y^2)\tau_T}{2}\right]}
    \label{eq:tunnelfilter}
\end{equation}
where $\tau_T=\frac{\sqrt{2I_P}}{\Efield}$ and $\Efield$ is the electric field strength.


The decisive influence of the tunnel filter function (\ref{eq:tunnelfilter}) makes it necessary
to eliminate it in order to retrieve a useful orbital image. This has been achieved in
Refs.~\cite{Meckel2008,Comtois2013} by directly comparing PMDs recorded for parallel and perpendicular
alignment, respectively. The difference images reveal clear structures that demonstrate that the PMDs
recorded with linear laser polarization contain a filtered projection of the orbitals.

Notably, circular polarization can also be used to map out the orbital shape in combination with
molecular orientation \cite{Akagi2009,Staudte2009,Holmegaard2010}. This approach, sometimes called
"laser STM" (as in scanning tunneling microscope) is similar to the attoclock. Here, however, the
unique mapping between the direction of tunneling and drift momentum is exploited to map out the
angle-dependence of the tunneling probability of aligned or oriented molecules. The partial Fourier
transform method \cite{Murray2010,Murray2011} explains how perpendicular PMD and angle-dependent tunnel
ionization yields are related to the orbital shape. The laser-STM technique has been utilized to resolve
angular correlations in sequential double ionization due to a spin-orbit wave packet in Neon cations \cite{Fleischer2011}.

\begin{figure}
    \centering
    \resizebox{0.5\textwidth}{!}{\includegraphics{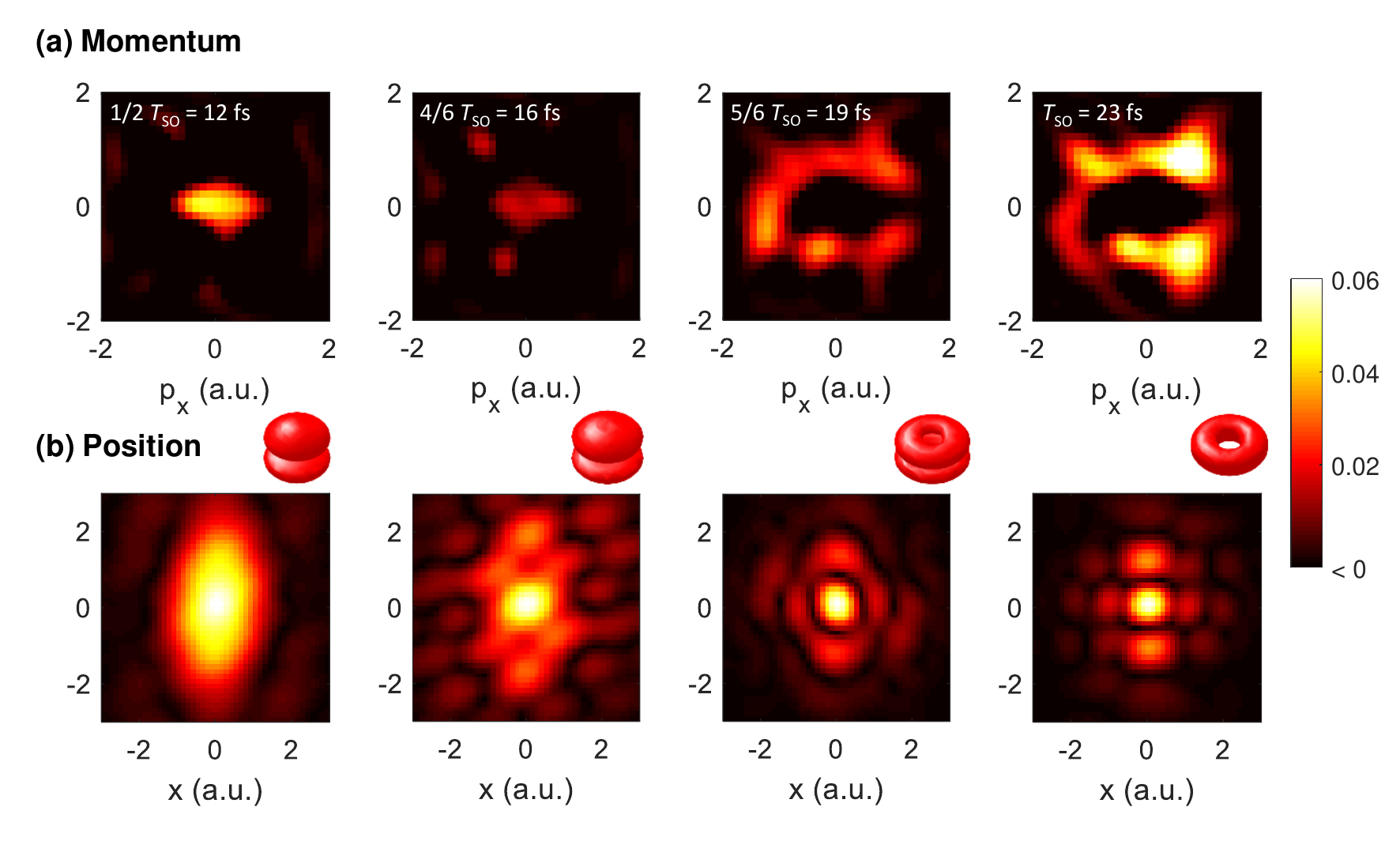}}
    \caption{Snapshots of an electron wave packet in the argon cation obtained by photoelectron orbital
    imaging. (a) Momentum-space images recorded at various fractions of the spin-orbit
    period $T_\mathrm{SO}$. The colorbar indicates the visibility of the delay-dependent
    variations in the recorded PMD. (b) Position-space images obtained by Fourier transform,
    assuming a flat phase in momentum space. The expected circular symmetry is broken by a
    stretch along $p_x$, due to an experimental artifact. The figure has been adapted from Ref.~\cite{kubel2019spatiotemporal}.}
    \label{fig:ar_so_wp}
\end{figure}

Direct imaging of a spin-orbit wave packet in Ar$^+$ has been recently achieved by combining
tailored laser fields \cite{kubel2019spatiotemporal}. The delay-dependent PMDs recorded in
coincidence with doubly charged ions deliver a movie of the electron motion shown in Fig.~\ref{fig:ar_so_wp}. In the
experiment, a few-cycle pump pulse is used to ionize neutral Ar, producing Ar$^+$ in a coherent superposition of
two spin-orbit states. This causes the vacancy in the Ar$^+$ valence shell to oscillate between
the $m = 0$ and $|m| = 1$ states with a period $T_\mathrm{SO} = 23.3\,\mathrm{fs}$, modulating the
spatial electron density. The momentum space signatures of these modulations are seen in the experimental
snapshots displayed in Fig.~\ref{fig:ar_so_wp}(a). After completion of a half-period $1/2\,T_\mathrm{SO}$,  the vacancy
is in the $|m|=1$ state, and the $m=0$ state, aligned with the laser polarization, is occupied by two electrons. At these delay
values, the measured electron density in momentum space exhibits a small spot in the center of the momentum distribution. For
alignment of the vacancy in the $|m|=0$ state, at $T_\mathrm{SO}$, a ring-shaped electron density is observed. The ring
shape can be understood as an image of the donut-shaped $|m|=1$ orbital, while the spot in the center relates to the peanut-shaped $m=0$ orbital.

Shown in Fig.~\ref{fig:ar_so_wp}(b) are the spatial images obtained by Fourier transform of the measured
momentum space distributions, assuming a flat phase. These spatial distributions do not correspond to
the actual spatial orbitals but rather to their autocorrelation signals. The discrepancy with respect
to the actual orbitals is most clearly seen for the donut-shaped distribution, which should not be
filled in the center. This illustrates how phase information is crucial to reconstruct the spatial orbitals.

\subsection{Phase information in electron momentum space}
\label{phase}

Unfortunately phases cannot be measured directly, and experimentally
only $|\Phi({\bf p})|^2$ is accessible (see chapter \ref{sec:wavepacketphase}). The phase of a
wave packet in final momentum space is relevant if this wave packet is superimposed with a second wave
packet which leads to interference. Here the absolute phase of the two wave packets does not change the
observable quantities and it is the relative phase of the two wave packets that determines if interference is
constructive or destructive. A few examples illustrating the relevance of relative phases in momentum space are briefly described below.

The interference of two electron wave packets that emerge from two different points in position
space, e.g. the atoms in a diatomic molecule, can act like a double-slit which gives rise to the
well-known interference pattern in momentum space for such a two-path
interference \cite{Akoury949,MaksimNatureCom}. As for a macroscopic double slit also here the
slit-geometry defines the observed interference pattern. 

ATI for a multi-cycle laser pulse leads to discrete values for the electron energy that can be explained
as a consequence of energy conservation or by an inter-cycle interference \cite{Arbo2010}. The time-dependent light field field
acts as a grating in the time-domain which is defined by the frequency of the photons. This gives rise to
interference in the energy-domain and the spacing of the peaks in the energy spectrum is proportional to the photon energy.

Sub-cycle interference occurs if two wave packets, which overlap in momentum space, are released at
times that differ by a timespan that is smaller than the duration of an oscillation of the light field. Conceptionally, sub-cycle
interference is very similar to inter-cycle interference \cite{Arbo2010,Eckart2018SubCycle,EckartArXivSideband}.  Examples that use
sub-cycle interference are two-color attoclock interferometry \cite{Han2018,Ge2019} and holographic angular streaking of
electrons \cite{Eckart2020HASETheo,DanielArXiv2020}. Interestingly such approaches allow for the access of changes of the Wigner time
delay in strong field ionization. The Wigner time delay is the derivative of the phase of the electron wave
packet with respect to energy \cite{Wigner1955,vos2018orientation}. However, the modeling of sub-cycle interference
by interference of electron wave packets that are born within less than one cycle of the laser field raises the question: Is it
possible to model sub-cycle interference (as in Refs. \cite{Arbo2010,Feng2019,EckartArXivSideband}) in the energy domain or is
the energy picture not suitable to model sub-cycle processes? Examples towards such a description are
found in Refs. \cite{Zipp_14,Kerbstadt_2017}. However, in order to calculate the coherent sum of all possible pathways in
the energy domain, the need for the inclusion of all the corresponding phases and amplitudes of these
pathways leads to a very high complexity.

Finally, laser-driven electron recollision leads to various types of interference and diffraction
effects, which we discuss in the following chapters \ref{sec:wavepacketphase} and \ref{sec:LIED}.

\section{Quantum interference of electron wave packets}
\label{sec:wavepacketphase}

The interference of two plane waves $\Psi_{1,2} = A_{1,2} \exp{\left(i\phi_{1,2}\right)}$, where $A$ is the complex amplitude and $\phi$ is the corresponding phase, yields
probability density
\begin{equation}
    \vert \Psi_1+\Psi_2 \vert^2 = \vert A_1 \vert^2+\vert A_2 \vert^2 + 2 A_1 A_2 \cos(\Delta \phi), \label{eq:interference}
\end{equation}
which allows access to the relative quantum phase $\Delta \phi = \phi_{1}-\phi_{2}$, and thus to
the natural space-time scales (atomic scales, \aa{}ngstr\"{o}m or nanometer and attosecond time) of electron dynamics.

In a next step, we review Quantum Spectral Phase Interferometry for Direct Electron wave-packet
Reconstruction (QSPIDER), which was introduced in Refs.~\cite{Remmetter2006,AChaconPRA2013}. In the experiment \cite{Remmetter2006}, EWPs are created by photoionization of atoms using an attosecond pulse train with a synchronized IR laser field, which induces a momentum shear to the EWP. Due to the periodicity of HHG, the sign of the momentum shear alternates between positive and negative for adjacent pulses. The resulting interference patterns allows for the reconstruction of the EWP's phase, and operates in close analogy to Spectral Phase Interferometry for Direct Electric-field Reconstruction (SPIDER)~\cite{Iaconis1998} in optical metrology.

\begin{figure}
    \centering
    \resizebox{0.5\textwidth}{!}{\includegraphics{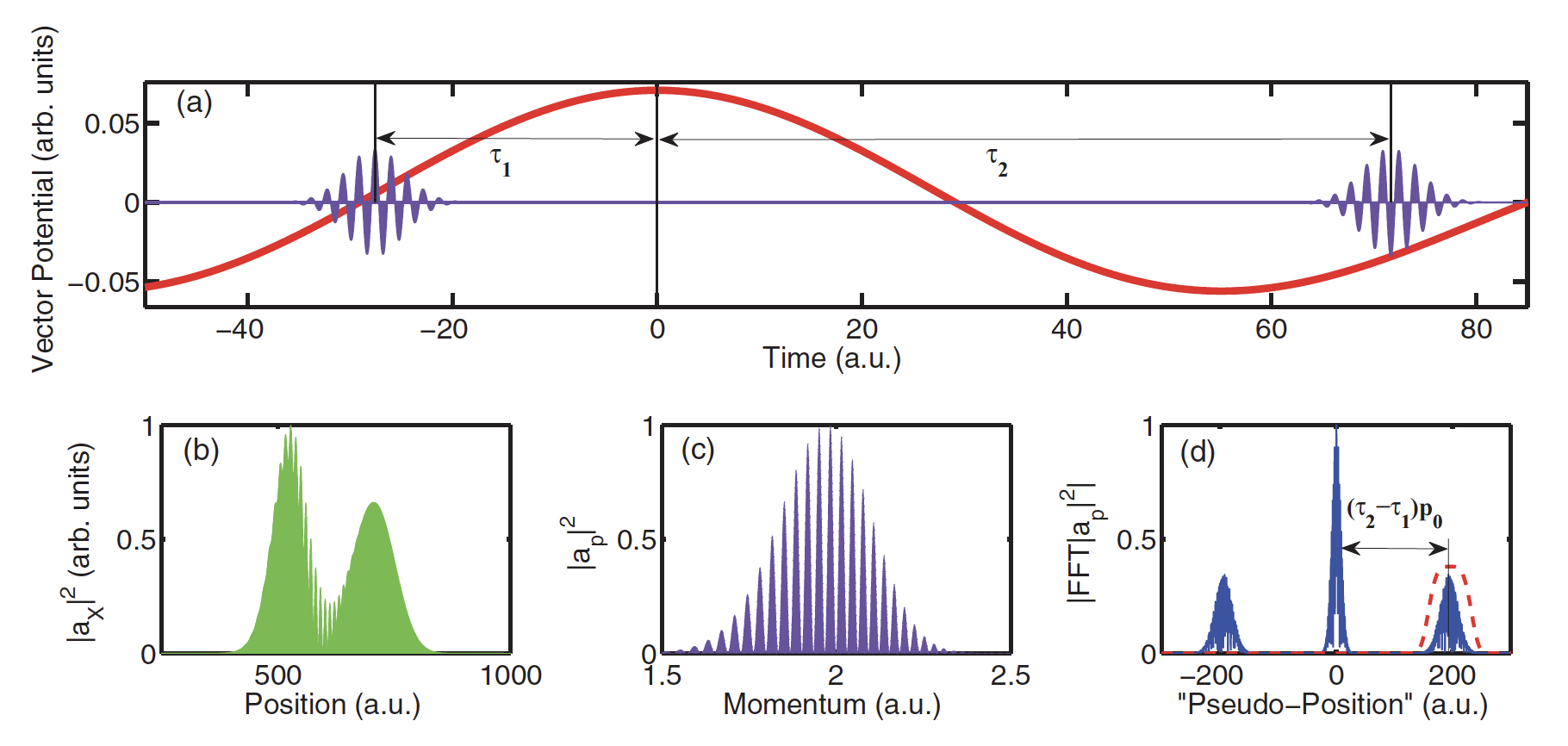}}
    \caption{Schematic representation of QSPIDER. (a) Depicts: (i) two
    identical XUV-attosecond pulses (in violet shadow), ${\bf E}_{\rm XUV,1/2}(t)$)
    with a delay $\tau_1$ and $\tau_2$ measured with respect to the zero time axis; (ii) In red,
    the vector potential ${\bf A}_L(t)$ of the electromagnetic IR laser-field with $800$~nm of
    wavelength. In (b), (c) we show the corresponding squared XUV-attosecond EWPs for real-space
    representation (in green) and momentum-space representation in (violet shadows), respectively. The momentum
    interference in (c) is the key in the extraction of the EWP phase difference. (d) Fourier decomposition
    of the momentum EWPs interference (c) which clearly has three peaks related to AC components
    centered at ${\tilde x}_{\pm}=(\tau_2-\tau_1)p_0$, obviously $p_0$ is the asymptotic EWP central momentum
    of the (c) and an DC component centered at ${\tilde x}=0$. Note, the pseudo-position space is indicated
    here by $\tilde x$. Figure reproduced from Ref.~\cite{AChaconPRA2013}}.
    \label{fig:QSPIDERFig}
\end{figure}
 Figure~(\ref{fig:QSPIDERFig}) shows the four steps towards the recuperation
 of a single EWP phase and its amplitude. In a first step, it is assumed that
 two identical EWPs are produced by replica of same extreme ultraviolet (XUV) attosecond pulse with
 the simple difference that there exists a time delay between them, as shown
 in Fig.~(\ref{fig:QSPIDERFig}a). The XUV-atom interaction at low
 intensities ($10^{10}-10^{12}~{\rm W}/{\rm cm}^2$) and for photon energies $\omega_{\rm X} > I_p$ exceeding the
 ionization potential, $I_p$, is described by perturbation theory. In the second QSPIDER step, a
 spectral shear is induced in the EWP by means of the weak infrared (IR) laser-field. 
 
In the third step, the final momentum-space distribution is
calculated. If the attosecond pulse duration is much shorter than
an optical cycle of the IR pulse, $T_0$, it can be expressed as the product
of an amplitude and, importantly, a phase factor. For an attosecond pulse centered at $\tau_1$ with respect
to the IR laser, the EWP is described by the following dependencies~\cite{AChaconPRA2013}:

\begin{align}
A({\bf p},t_{\rm F},\tau_1) \propto& \, |{ d}[{\bf p} + {\bf A}_L(\tau_1) ](t_{\rm F})| \label{eq:qEWP00} \\
\phi({\bf p},t_{\rm F},\tau_1)  \propto&\, \varphi_{d[{\bf p}+{\bf A}_L(\tau_1)]}(t_{\rm F}) \label{eq:qEWP01}
\end{align}

\noindent where ${d}({\bf p}) = {\bf e}_x\cdot {\bf d}({\bf p})$
is the $x$-direction component of the dipole transition matrix element
${d}({\bf p})=\langle {\bf p}\left| (-{x})\right| \Psi_0 \rangle$ along the common
polarization direction of both the IR laser field and XUV-attosecond pulse,
$|\Psi_0\rangle$ is the ground state of the atomic or molecular system, $|{\bf p}\rangle$ is the
scattering continuum wave and $-{x}$ is the dipole moment operator which is proportional to the position operator ${x}$.
Thus, the physical interpretation of $d({\bf p})$ is the complex transition from the ground state  $|\Psi_0\rangle$ to
the continuum state  $|{\bf p}\rangle$ mediated by the dipole operator ${(-x)}$~\cite{SakuraiBook}.
In case of a single attosecond EWP, the amplitude in Eq.~(\ref{eq:qEWP00}) $A({\bf p},t_{\rm F},\tau_1)$ is
proportional to the real amplitude of the dipole transition matrix element $|d({\bf p})|$. In Eq.~(\ref{eq:qEWP01}), the dipole
phase is $\varphi_{d[{\bf p}+{\bf A}_L(\tau_1)]}$ which will be extracted as in Ref.~\cite{AChaconPRA2013}.
In general, the laser-induced chirp (LIC) generated by
the variations of the IR field around the time of ionization
needs to be considered. However, in the case of short attosecond pulses ($<$200 as) and
modest intensity ($I_0$~$<$~$10^{13}$~W/cm$^2$) the effects of the LIC phase are negligible. This phase depends
on the value of the electric field at ionization time $\tau_1$ and is zero if ${\bf E}_L(\tau_1) = {\bf 0}$~\cite{AChaconPRA2013}. It will
become relevant for streaking and interferometric measurements as it can become larger than the phase of the dipole.

The most important aspect in this third step is to recover the dipole phase
difference $\Delta \varphi_{d[{\bf p}+{\bf A}_L(\tau_1,\tau_2)]}$ which can be approximated as
\begin{align}
    \Delta \varphi_{d[{\bf p}+{\bf A}_L(\tau_1,\tau_2)]} \approx \frac{\partial \varphi_d(p)}{\partial p}\Delta A_L \label{eq:RPhase}
\end{align}

The last step in QSPIDER is to apply the Fourier algorithm to extract the derivative of the dipole phase and integrate it. In the next section, we
will follow those steps in He$^+$.

\subsection{Quantum spectral phase interferometry for direct electron wave-packet reconstruction}
\label{qspidersection}

The validity of the QSPIDER concept has been
verified by a numerical simulation presented in Ref.~\cite{AChaconPRA2013}. To this end,
two delayed copies of an EWP with a relative shear between them are used to construct an
interferogram, similar to the optical SPIDER technique. This can be realized by focusing
an attosecond pulse train (APT) with exactly two pulses centered at $\tau_1$ and $\tau_2$, onto He$^+$ in the
presence of a weak IR laser pulse with vector potential ${\bf A}_L(t)$. One obtains two EWPs which are
delayed relative to each other by approximately one optical cycle of the IR-laser. The IR laser streaks
each of the EWPs resulting in a relative streaking, $\Delta{\bf A}_L = {\bf A}_L(\tau_2) - {\bf A}_L(\tau_1)$, between the two
EWPs copies. The streaked and delayed copies produce an interferogram in the final momentum distribution which is conceptually
equivalent to the interferogram of the SPIDER technique (see Fig.~\ref{fig:QSPIDERFig}c) ~\cite{AChaconPRA2013}.

By applying the 4-steps described in the previous section and the Fourier analysis (see Fig.~6d), we can extract the interesting dipole matrix phase of Eq.~(\ref{eq:RPhase}), which in certain limit is the momentum derivative of the dipole phase. We can also extract the EWP  $\left| A({\bf p}) \right|$ associated to the DC term in the limit ${\bf A}_{L}(\tau_1) \approx {\bf A}_{L}(\tau_2)$. In the next section, we will apply QSPIDER principles to He$^+$ and demonstrate the retrieval of the Dipole matrix elements.

\subsection{Quantum spectral phase interferometry for direct electron wave-packet reconstruction in 2p states}

\begin{figure}
    \centering
    \resizebox{0.5\textwidth}{!}{\includegraphics{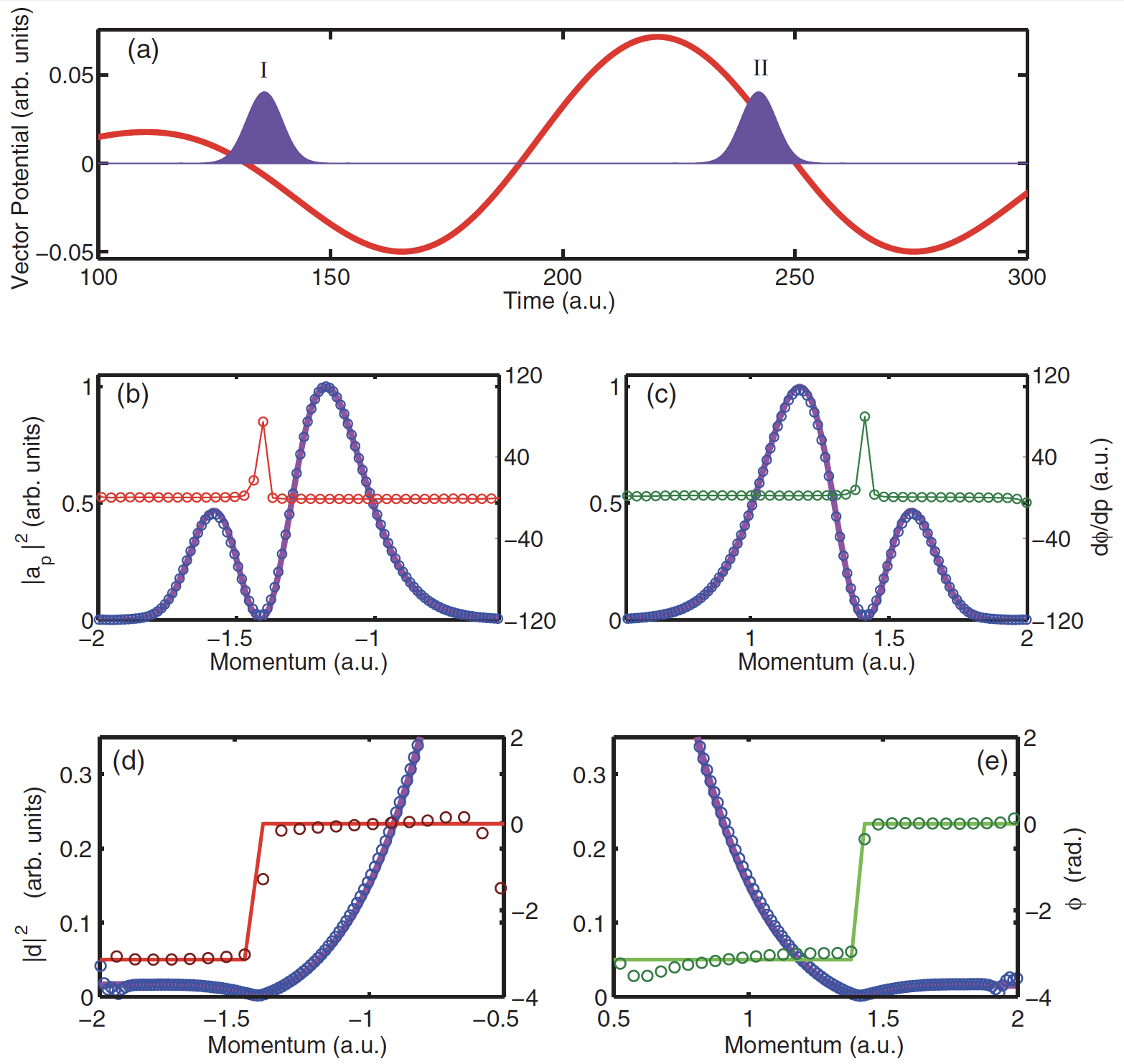}}
    \caption{QSPIDER retrieval for the first excited state of the He$^+$  based on strong-field approximation momentum distributions. (a) The vector potential of the IR laser pulse in red line. The field envelope of the APT in violet areas. The retrieved momentum distribution of the EWP (blue circles) compared to the exact momentum distribution (violet solid line) from the interaction with a single attosecond pulse without IR field is plotted in panels (b) and (c) for $p < 0, \,\&, \, p > 0$. The retrieved derivative of the dipole phase is shown in red and green circles, respectively, in panels (b) and (c). The retrieved squared amplitude (blue circles) and phase (red and green circles) of the dipole matrix element is compared to the exact squared amplitude (violet solid line) and phase (red and green solid lines) in panels (d) and (e).}
    \label{fig:QSPIDERFigHe}
\end{figure}

\noindent The two XUV-ATP in the presence of a weak IR laser pulse interacting with
He$^+$ are shown in Fig.~\ref{fig:QSPIDERFigHe}. This configuration creates an ideal
interferometry scenario to extract the dipole phase derivative and the amplitude.

By applying the Fourier Analysis of SPIDER to the AC component, the extraction of
this dipole phase derivative is shown in Fig.~\ref{fig:QSPIDERFigHe}(b) and~\ref{fig:QSPIDERFigHe}(c) for
negative and positive p-momentum up to the spectral range which the XUV-pulse allow us. The EWP amplitude
has a clear node at $p\sim \pm 1.5$~a.u., as expected for 2p orbitals. The red and green
dots show the QSPIDER reconstruction of the dipole matrix element derivative, which was obtained
by dividing the EWP amplitude by the XUV-spectral amplitude (see Eq.~(\ref{eq:qEWP00}) for the
EWP amplitude $A({\bf p}, t_{\rm F},\tau_1)$). Concerning the dipole phase reconstruction, we
observe clearly a Dirac-like distribution, unique for a system in which the phase has a jump of $\pi$. This jump
is shown in Fig.~\ref{fig:QSPIDERFigHe}(d) and (e) for positive an negative momenta, and compared to the analytic
dipole phase. Good agreement between the reconstruction and the expected dipole phase is found. We also
performed TDSE calculation in 1D which are detailed in Ref.~\cite{AChaconPRA2013}.

The example of QSPIDER demonstrates how quantum interference of EWPs provides access to phase information of
the EWPs, and by extension, of the atomic or molecular system under study. While the first EWP propagates
in the continuum, the second one remains bound until the second attosecond pulse arrives, meanwhile
probing the system. The information carried by the second EWP is retrieved by considering the first EWP as
a known reference wave. This is the concept of holography, and it is applicable to a larger number
of experiments in strong-field and attosecond physics, in particular to electron rescattering in ATI.

\subsection{Photoelectron holography and its limitations}
\label{sec:holography}

\begin{figure}[t]
    \centering
    \resizebox{0.5\textwidth}{!}{\includegraphics{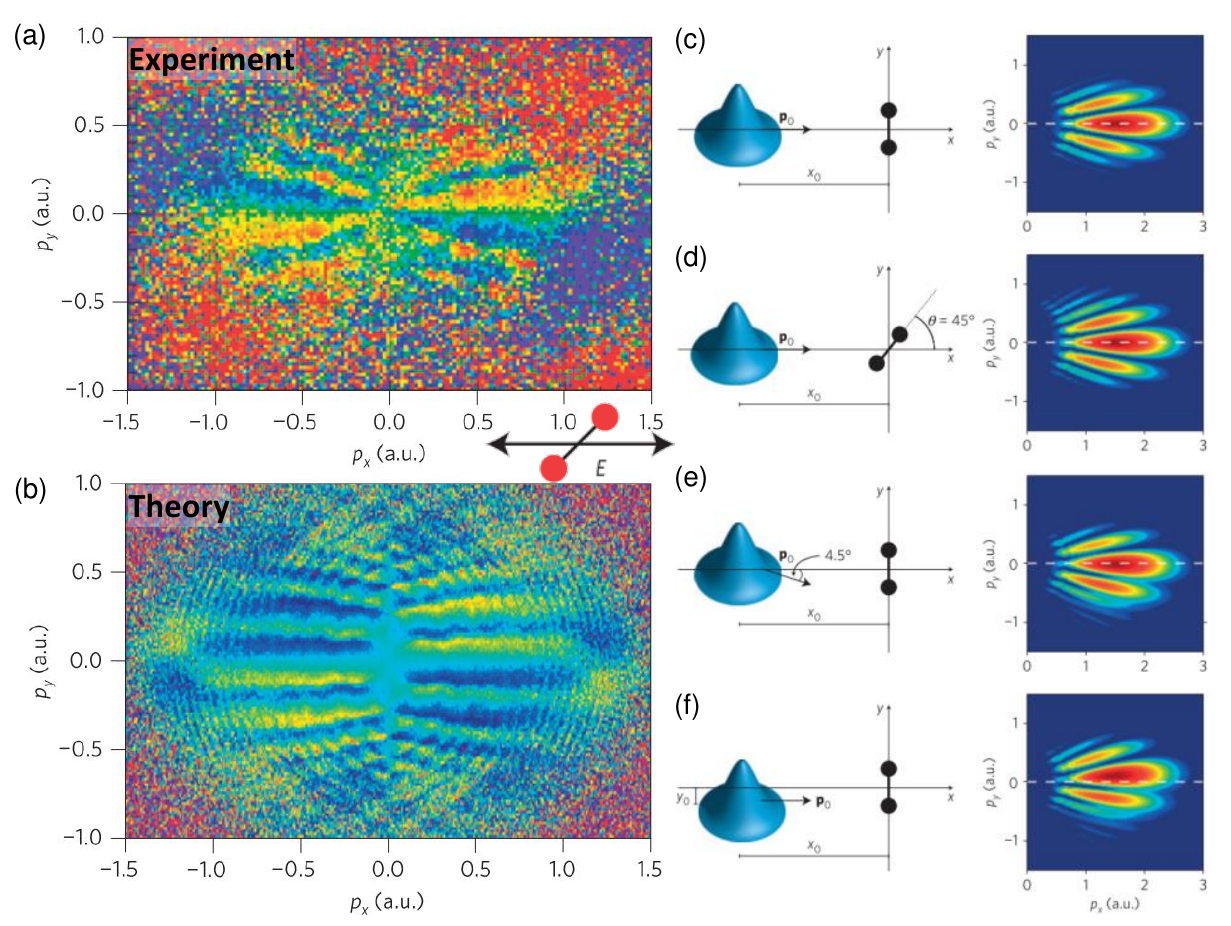}}
    \caption{Photoelectron holography experiment with aligned N$_2$. As indicated by the cartoon in the center of the plot, the data shown in (a)
    is the normalized difference of the PMD recorded for an alignment angle of $45^\circ$ and the PMD
    integrated over all alignment angles. The
    important feature is the off-center holographic fringe pattern, that is qualitatively
    reproduced by the theoretical results shown in (b).    Panels (c-f) show results of
    wavepacket scattering simulations for different scenarios: (c) symmetric
    reference scenario, (d) two-center potential set to an angle of $45^\circ$, (e) incident Gaussian
    wavepacket launched at an angle of $-4.5^\circ$, (f) incident Gaussian wave packet offset in
    the vertical direction ($y_0=-4\,\au$). The color scale is logarithmic and covers two orders of magnitude. Figure adapted
    from Ref.~\cite{Meckel2014} with permission from Nature Publishing Group.}
    \label{fig:meckel2014}
\end{figure}

When the liberated electron in ATI is driven back to the core, it scatters on the (ionic) potential. In the
simplest case of a point-like scatterer, the scattered wave can be approximated as a spherical wave originating
at the core. If we consider the unscattered wave as a plane wave and interfere it with the scattered
spherical wave, we obtain an interference pattern similar to the one shown in Fig.~\ref{fig:meckel2014}(c). It closely resembles the
well-known side lobes in the angular distribution of of ATI first reported in Ref.~\cite{Yang1993}, which has been known
as holographic interference pattern since the landmark papers from Spanner et al. and Huismans et al.~\cite{Spanner2004,Huismans2011}. The
holographic interpretation of these features allows one to utilize them as a probe of the scattering potential and associated
dynamics. Variations in holographic patterns have been used to probe ionization
dynamics in two-color experiments~\cite{Porat2018,Eckart2018SubCycle,EckartArXivSideband,Han2018,Ge2019,Skruszewicz2015}, molecular
dissociation~\cite{Haertelt2016}, and bound electron and nuclear dynamics~\cite{Walt2017}.

However, a word of warning comes from an important paper by Meckel, \textit{et al.}, who studied
the effect of molecular alignment on the holographic fringes in PMD~\cite{Meckel2014}.
Extending on their pioneering work on LIED~\cite{Meckel2008}, the authors carefully varied the angle
between the molecular axis and the laser polarization, and found a striking off-center holographic
fringe pattern for an angle of $45^\circ$ that is reproduced in Fig \ref{fig:meckel2014} (a). This
pattern agrees well with results obtained by numerically solving the TDSE, Fig.~\ref{fig:meckel2014}(b). For the interpretation
of these results, and to understand the meaning of the off-center fringe pattern, simple wave packet scattering
simulations are used. These demonstrate that it is not the tilt of the molecular axis that moves
the fringes (Fig.~\ref{fig:meckel2014}(c, d). It is rather a property of the recolliding wave packet
that explains the observations. Specifically, if the wave packet is given a spatial offset
relative to the molecular axis, the off-center
fringe pattern is obtained [Fig.~\ref{fig:meckel2014}(f)]. This led the authors to
conclude that in their ``experiment, electron holography provides information about
the continuum electron wave packet rather than the scattering object"~\cite{Meckel2014}. This study
demonstrates that it is important to know the relevant properties of the recolliding electron wave
packet in order to adequately probe molecular structure.

Nevertheless, the example of QSPIDER from section \ref{qspidersection} shows that recollisions
are not a necessary prerequisite for holography. An example that harnesses photoelectron
holography without recollision is holographic angular streaking of electrons (HASE)~\cite{Eckart2020HASETheo} where a
co-rotating two-color laser field is used to create two electron wave
packets that interfere and reveal properties of the phase of the electron
wave packet in momentum space. Since for HASE the combined electric field
is close to circularly polarized, the continuum wave function, the wave
function at the tunnel exit and the bound electron wave function are
closely related~\cite{DanielArXiv2020}. This is in contrast to linearly
polarized light where recollision and sub-cycle interference lead to a non-trivial
relationship of the continuum wave function and the wave
function at the tunnel exit~\cite{PhysRevA.102.013109,Figueira_de_Morisson_Faria_2020}.

\section{Laser-induced electron diffraction}
\label{sec:LIED}
The process of electron rescattering can lead to interference, even in the
absence of an unscattered reference wave. This phenomenon is known
as Laser-induced electron diffraction (LIED)~\cite{Corkum2007,Amini2020,Lein2002,Meckel2008,Blaga2012,Zuo1996,Lin2010,Okunishi2011,Xu2012,Xu2014} and is
the strong-field-variant of ultrafast electron diffraction (UED) whereby a molecule is tunnel ionized
to generate an EWP that is used to take a ``selfie'' of its molecular structure.
LIED can retrieve the internuclear distances in a molecule with picometer
and
attosecond
precision.

LIED’s extension to the more advantageous mid-infrared (MIR; i.e. $\lambda \gtrsim \,2\,\mu$m) wavelength range
has enabled the direct retrieval of many diatomic and more complex molecular structures~\cite{Amini2020,Wolter2016,Wolter2015,Pullen2015,Pullen2016,Amini2019a,Amini2019b,Ueda2019,Fuest2019,Liu2019,Belsa2020,Sanchez2020,Liu2021}.
In fact, not only is the de Broglie wavelength $\lambda_B$ of the rescattering electron
significantly smaller for a driving laser at $\lambda = 3\,\mu$m ($\lambda_B \sim 0.75\,\AA$) than
at $\lambda = 0.8\,\mu$m ($\lambda_B \sim 2.75\,$\AA).
The longer driving wavelength also yields a relatively large lateral EWP extent of $\Delta x>\,150\,\AA$
relative to $\Delta x <\,50\,\AA$ at $\lambda = 0.8\,\mu$m~\cite{Amini2020,Wolter2015,Ueda2019}.

Notably, these properties have enabled LIED in the MIR to capture a sub-10-fs snapshot
of deprotonation in dissociating C$_2$H$_2^{2+}$. This was only possible with LIED’s
sub-optical-cycle probe of molecular structure together with its sensitivity to hydrogen
scattering. Moreover, ultrafast changes on the rising edge of the LIED pulse have been shown to
lead to significant structural deformation in C$_{60}$~\cite{Fuest2019}, CS$_2$~\cite{Amini2019a}
and OCS~\cite{Sanchez2020}.

LIED can be well-described using the laser-driven electron-recollision framework
\cite{Corkum1993,Kulander1993,Schafer1993,Varro1993} in which the emitted EWP is: (i) accelerated by
the oscillating electric field of the intense laser pulse before (ii) returning and (iii)
rescattering against the target ion. It is justified to consider only the dominant trajectory
that leads to a given drift momentum after rescattering. In the case of LIED this is the so-called
long trajectory, which is produced close to the peak of the electric field. In the quantum mechanical
picture of LIED shown in Fig.\,\ref{fig1}, the emitted EWP is returned and rescattered against two
scattering centers (i.e. two atoms in a molecule), leading to interference fringes in the detected
electron momentum distribution. These interference fringes are described by the coherent molecular
interference term, $I_{\rm M}$, which  contains structural information and can be in the framework
of the independent atom model (IAM)~\cite{Lin2010,Schafer1976,McCaffrey2008}, as
given by~\cite{Amini2020,Hargittai1988}
\begin{equation}
    I_{\rm M}(q)\propto \sum_{i=1}^{N} \sum_{j=1}^{N} f_{\rm i}(q)f_{\rm j}^*(q)e^{i({\bf q}\cdot {\bf R_{\rm ij})}},
\end{equation}
Specifically, the phase factor of $I_{\rm M}$ contains the internuclear distance between
two atoms ($i$ and $j$), $R_{\rm ij}$. $I_{\rm M}$ is a function of the
momentum transfer (i.e. $q = 2k_r\,\times\,\sin(\theta_{\rm r}/2)$) between
the incoming EWP and target following scattering, where $f_{\rm i}$ is the electron
scattering amplitude on atom $i$. In fact, $I_{\rm M}$ is detected as a
sinusoidal signal for randomly oriented molecules as given by~\cite{Amini2020,Hargittai1988}
\begin{equation}
    I_{\rm M}(q)\propto \sum_{i=1}^{N} \sum_{j=1}^{N} f_{\rm i}(q)f_{\rm j}^*(q)(\frac{\sin(qR_{\rm ij})}{qR_{\rm ij}}),
\end{equation}
which typically appears as oscillations in the detected momentum distribution
of high-energy electrons.

\begin{figure}[t!]
 \centering
 \resizebox{0.49\textwidth}{!}{
  \includegraphics{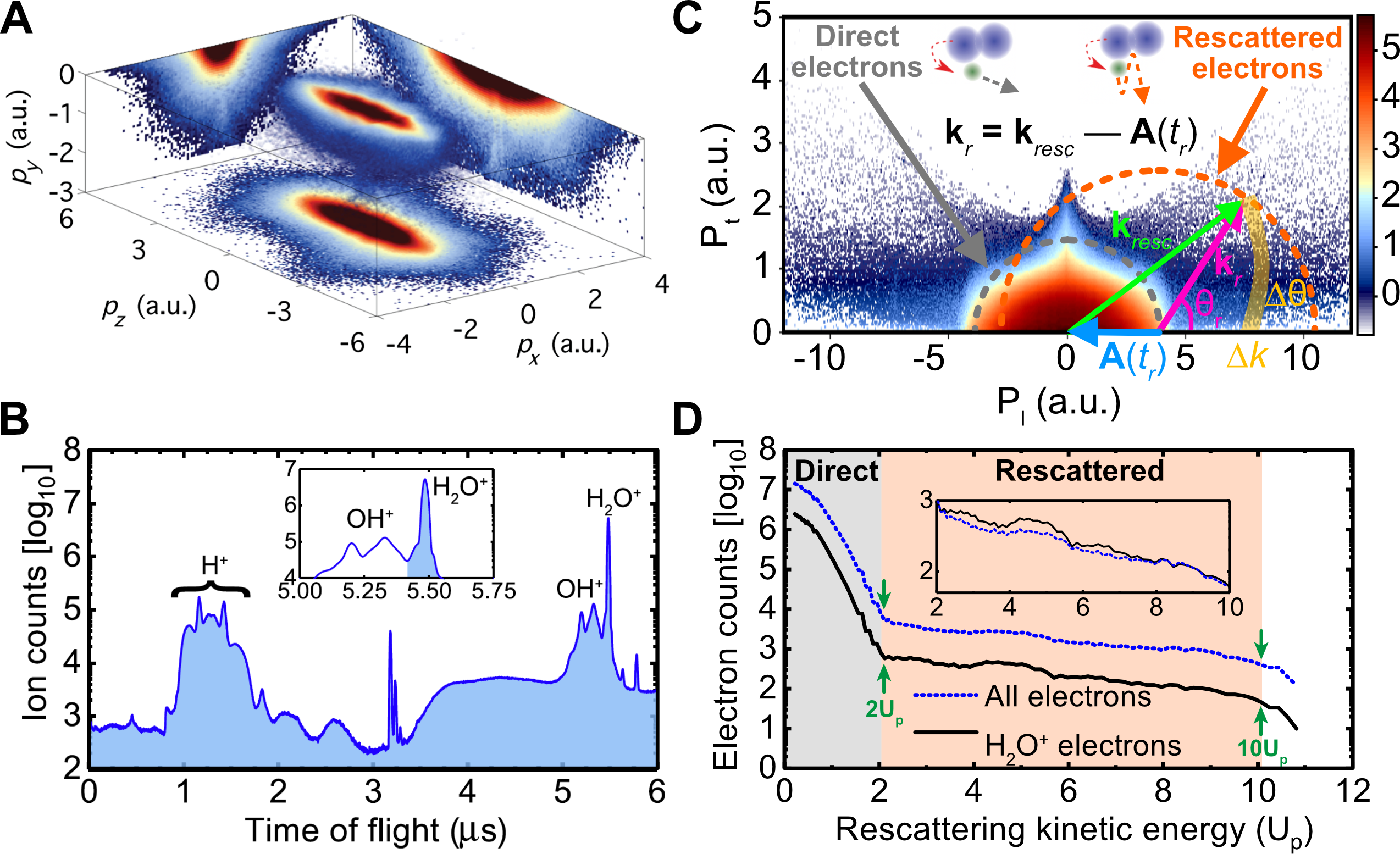}}
  \caption{(\textbf{LIED rescattered electron data measured with a reaction microscope (ReMi). (a)} Three-dimensional electron
  data measured with a reaction microscope.~\cite{Wolter2015}. \textbf{(b)} Typical ion time-of-flight (ToF) spectrum
  measured from the LIED of H$_2$O. The inset shows the ToF range corresponding to the most dominant
  ToF peak, H$_2$O$^+$, shaded in blue. \textbf{(c)} Two-dimensional longitudinal-transverse electron
  momentum distribution (i.e. $P_\parallel$-$P_\bot$). A schematic of typical direct (grey dashed) and
  rescattered (orange dashed) electron momentum distributions is shown. A diagram defining the
  relationship between the return momentum, $k_{\rm r}$, and the vector potential, $A(t_{\rm r})$, at the
  instance of rescattering, $t_{\rm r}$, with the detected rescattered momentum, $k_{\rm resc}$. The DCS is
  extracted by integrating the electron signal in the area given by the yellow block arc composed of
  $\Delta\theta$ and $\Delta\,k$. \textbf{(d)} Electron counts as a function of rescattering kinetic
  energy in units of $U_{\rm p}$ for all electrons (grey shaded) and H$_2$O$^+$ electrons (orange shaded).
  The $2U_{\rm p}$ and $10U_{\rm p}$ classical cut-offs are indicated by green arrows. The direct and
  rescattered regions are indicated by grey and orange shaded areas, respectively. Inset shows zoom-in of the
  rescattered region. Figures adapted from Ref.~\cite{Wolter2015,Liu2019}. }
  \label{fig:1kas}
 \end{figure}

The full three-dimensional momentum distribution of rescattered electrons, as
shown in Fig.\,\ref{fig:1kas}(a), can be detected, for example, with a
COLTRIMS reaction
microscope (ReMi)~\cite{jagutzki2002multiple,ullrich2003recoil,Moshammer1996,Dorner2000} (also see section \ref{section_coltrims}). Importantly, the ReMi can simultaneously detect electrons and ions in kinematic coincidence to select the electron-ion fragmentation channel that is generated during the intense-laser matter interaction in a two-step process. Firstly, the ion of interest (e.g. H$_2$O$^+$)~\cite{Liu2019} is identified by selecting its corresponding ion time-of-flight (ToF) range from the ion ToF spectrum, see Fig.\,\ref{fig:1kas}(b). Then the two-dimensional electron momentum distribution parallel, $p_{\parallel}$, and perpendicular $p_{\bot}$ of electrons generated with the ion of interest can be generated,  see Fig.\,\ref{fig:1kas}(c). Here, the return momentum, $k_{\rm r}$, at the time of scattering, $t_{\rm r}$, is obtained by subtracting the vector potential, $A(t_{\rm r})$, of the laser field from the detected rescattered momentum, $k_{\rm resc}$. The differential cross-section (DCS; i.e. number of electrons scattered into a specific solid angle) is extracted by integrating the block arc (yellow) area in Fig.\,\ref{fig:1kas}(c) at various different $k_{\rm r}$.

Fig.\,\ref{fig:1kas}(d) shows the measured electron yield from ionization of H$_2$O~\cite{Liu2019} for all electrons (blue dashed) and
electrons detected in coincidence with H$_2$O$^+$ (black solid) as a function of kinetic
energy in units of ponderomotive energy ($U_{\rm p}$; i.e. the cycle-averaged kinetic energy of a
free electron oscillating in the electric field of a laser pulse). Fig.\,\ref{fig:1kas}(d) has two
regions clearly distinguishable. Electrons that rescatter (do not rescatter) against the target ion are
detected with a typical rescattering kinetic energy of  $2\,-\,10\,U_{\rm p}$ ($0\,-\,2\,U_{\rm p}$) and are
referred to as “rescattered” (“direct”) electrons as indicated by the orange (grey) shaded
regions in Fig.\,\ref{fig:1kas}(d). Thus, as a second step, only the $2\,-\,10\,U_{\rm p}$ rescattered
region is considered for LIED imaging. In this region, one can see that the sinusoidal
signal of the $I_{\rm M}$ is more pronounced in the H$_2$O$^+$-electron data as
compared to the all-electron data. This demonstrates the capability of electron-ion
coincidence detection with a ReMi to provide a more sensitive probe of the $I_{\rm M}$ which would otherwise
be washed out by background signal without coincidence selection.

\begin{figure}[t!]
 \centering
 \resizebox{0.49\textwidth}{!}{
  \includegraphics{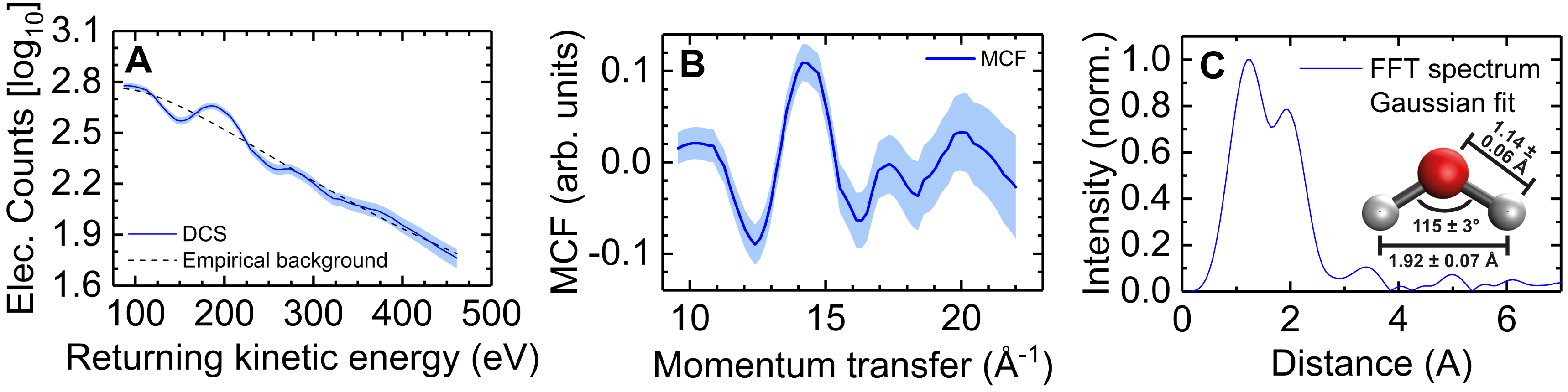}}
  \caption{\textbf{FT-LIED retrieval of H$_2$O$^+$ molecular structure. (a)} DCS (blue solid) and
  empirical background (black dashed) as a function of returning kinetic energy. Uncertainty in
  DCS is shown as blue shaded area. \textbf{(b)} MCF as a function of momentum transfer. \textbf{(c)} Radial
  distribution following the Fourier transform of panel (B). Figure adapted from Ref.~\cite{Liu2019}.}
  \label{fig:2kas}
 \end{figure}

In fact, highly-energetic electrons with detected kinetic energies of
hundreds-of-eV range (i.e. $U_{\rm p} \gg 10\,$eV, see Fig.\,\ref{fig:2kas}(a)) are
required to achieve an appreciable momentum transfer, $q$, to penetrate beyond the
valence electron cloud and scatter against the inner-most core electron shell close
to the nuclei. To achieve the high kinetic electron energies, long wavelength driver
sources (i.e. $\lambda>\,2\,\mu$m in the mid-infrared range) are needed to drive
LIED experiments. Extracting the molecular interference signal $I_{\rm M}$ from the total
interference signal $I_{\rm T}$ requires the subtraction of the background atomic $I_{\rm A}$ signal. This can be
achieved by either calculating the $I_{\rm A}$ signal using the IAM or by applying a background empirical
fit to the detected DCS signal, the latter of which is shown in Fig.\,\ref{fig:2kas}(a). Doing so
allows the $I_{\rm M}$ molecular signal to be contrasted against the $I_{\rm A}$ atomic signal
through the molecular contrast factor, MCF, as given by~\cite{Amini2020,Hargittai1988}
\begin{equation}
    \mathrm{MCF} = \frac{I_{\rm T}-I_{\rm A}}{I_A} = \frac{I_{\rm M}}{I_{\rm A}}.
\end{equation}
Fig.\,\ref{fig:2kas}(b) shows the MCF as a
function of momentum transfer, which provides a unique
fingerprint of the molecular structure through the sinusoidal
signal that is related to the $I_{\rm M}$. Fourier-transforming (FT) the MCF signal provides the
one-dimensional radial distribution of internuclear distances that are present in the molecule. In this
case for the LIED imaging of H$_2$O, two FT peaks are clearly present that correspond to the O-H and H-H internuclear
distances at 1.14 and 1.92\,$\AA$ when comparing to literature values.~\cite{Liu2019}

\begin{figure}[t!]
 \centering
 \resizebox{0.49\textwidth}{!}{
  \includegraphics{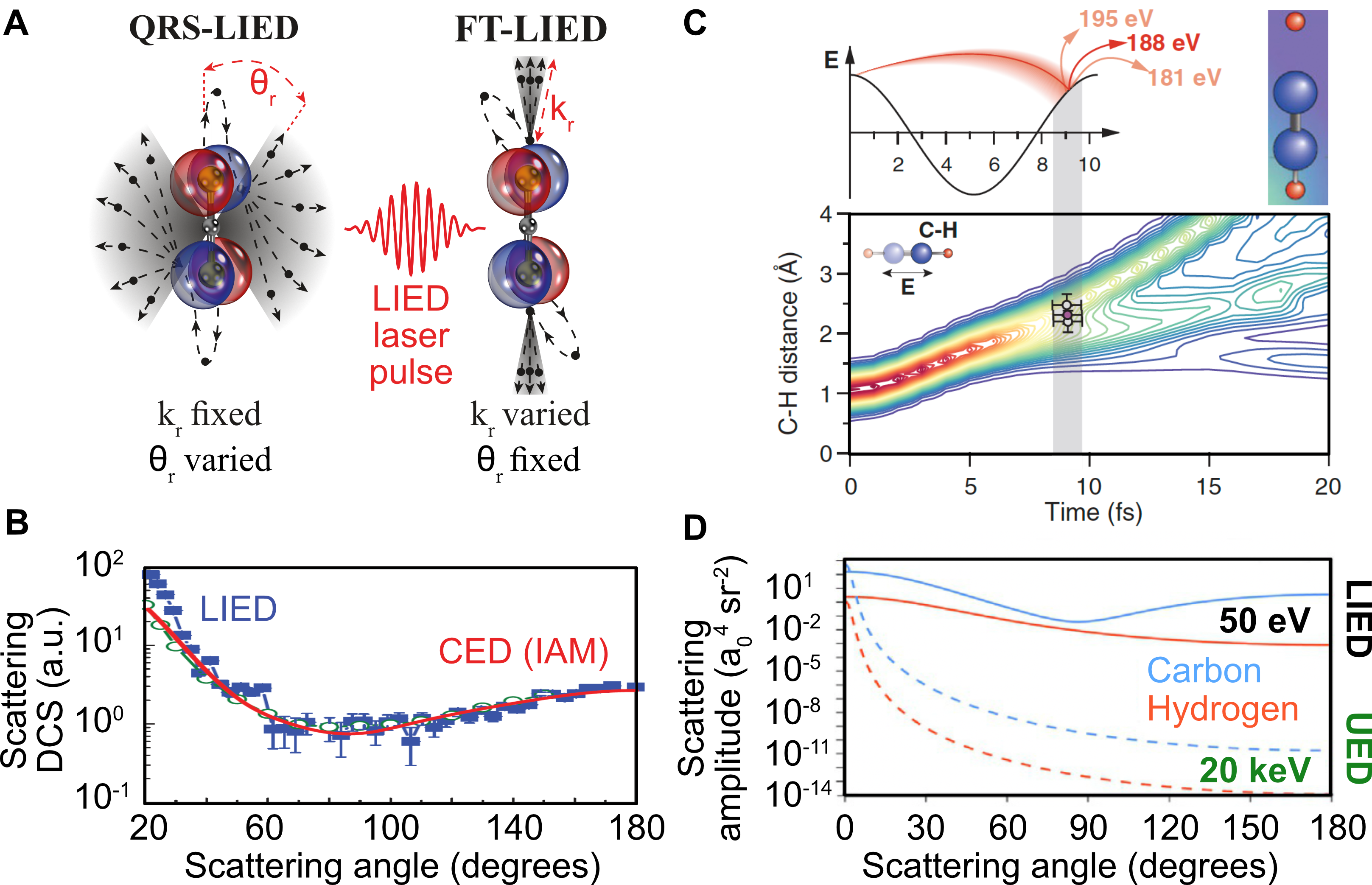}}
  \caption{\textbf{Variants and characteristics of LIED. (a)} Schematic of QRS-LIED and FT-LIED.
  \textbf{(b)} Scattering DCS as a function of scattering angle. LIED (blue squares) and CED
  data (red line) are shown. \textbf{(c)} Extracted C-H internuclear distance as a function of
  time after ionization following the LIED imaging of C$_2$H$_2^{2+}$. \textbf{(d)} Scattering
  amplitude as a function of scattering angle for carbon (blue) and hydrogen scattering (orange)
  at typical LIED (solid) and UED (dashed) electron kinetic energies of 50 eV and 20 keV,
  respectively. Figures adapted from Refs.~\cite{Blaga2012,Wolter2015,Wolter2016,Ueda2019}.}
  \label{fig:3kas}
 \end{figure}

There in fact exist two variants of LIED, as shown in Fig.\,\ref{fig:3kas}(a): (i) FT-LIED~\cite{Pullen2015,Amini2020} or also
called fixed-angle broadband laser-driven electron scattering (FABLES)~\cite{Xu2014}, and (ii) LIED based on the
quantitative rescattering (QRS) model~\cite{Chen2009,Lin2010}, referred to as QRS-LIED.
In FT-LIED, the energy dependence of rescattering in back-rescattered electrons are
only considered (i.e. varied $k_r$ at fixed $\theta_r\,\approx\,180^\circ$). Here, the far-field detected
electron momentum distribution can be related to the near-field image of the molecular structure through a FT relation. In QRS-LIED, only the angular dependence of rescattering (i.e. varied $\theta_r$ at fixed $k_r$) is considered at various fixed $k_r$ enabling the measurement of the doubly differential cross-section (DDCS) of elastic scattering. Fig.\,\ref{fig:3kas}B shows the angular-dependence of scattering in N$_2$ measured with LIED (blue squares) and with field-free conventional electron diffraction (CED; red line).~\cite{Blaga2012} The very good agreement between LIED and CED demonstrates LIED’s ability to extract field-free DCS from field-dressed measurements which are comparable to those measured with CED. Moreover, LIED’s sensitivity to hydrogen scattering is demonstrated by its structural retrieval of many hy\-dro\-gen-con\-tain\-ing molecules such as in the deprotonation of C$_2$H$_2^{2+}$ (see Fig.\,\ref{fig:3kas}C), as well as in C$_2$H$_2$, H$_2$O, NH$_3$, C$_6$H$_6$ and more~\cite{Amini2020}. This is particularly pronounced at scattering angles other than forward scattering (i.e. $\theta_r\,>\,10^\circ$) where the scattering amplitude of hydrogen scattering is within an order of magnitude of carbon scattering in LIED due to the low kinetic energies of the rescattering LIED electron (see Fig.\,\ref{fig:3kas}D).
Whilst in UED, the scattering amplitude of hydrogen and carbon scattering at $\theta_r\,>\,10^\circ$ is orders-of-magnitudes lower
than in LIED owing to the significantly higher electron kinetic energies used in UED. Although UED is limited to
forward-scattering-only, time-resolved UED studies have demonstrated to be a very sensitive and powerful
probe of molecular structure and photoinduced molecular dynamics
using sub-150-fs MeV UED electron pulses~\cite{Amini2020,Yang2018,Wolf2019,Yang2020}.
A variety of complementary aspects between field-dressed LIED and field-free UED measurements exist,
with many future opportunities to study a variety of gas-phase molecular structures and associated
dynamics~\cite{Amini2020}.

\section{How to extract photoelectron momentum distribution from the ab initio calculations?}
\label{sec:benjamin}

Over the last three decades numerical solutions of the time-dependent Schr\"{o}dinger equation 
within the single-active electron (SAE) approximation have emerged as one of the main theoretical tools used
to study photoionization and strong-field phenomena. Due to the ``black-box'' nature of the 
TDSE, however, the underlying physics are often
interpreted using alternative theoretical models \cite{Armstrong2021} and approaches such as
 the strong-field approximation
(SFA, for a review see Refs.~\cite{Becker2002,miloTP}). Other approaches, which are often variants of the SFA, include quantum-orbit
theory~\cite{QO2000,kopprl,science,JMO}, 
Coulomb-corrected SFA~\cite{ccsfa,tccsfa}, Coulomb quantum-orbit strong field approximation~\cite{lai,cqsfa,cqsfa1},
semiclassical two-step model~\cite{dimitrovski,scts,scts1}, classical trajectory based Monte-Carlo method~\cite{ctmc,hofman,ctmc1},
quantum trajectory based Monte-Carlo method~\cite{qtmc} and many more.

The majority of these methods have one aspect in common, namely, they use a trajectory
based picture to describe the field-induced ionization process and the associated electron motion.
Importantly, there may exist many different pathways for an electron to reach the
detector with the same final linear momentum.
Trajectory based methods allow us to explain features in the photoelectron
momentum distribution as quantum interference of different pathways, yielding an
intuitive physical interpretation of the photoionization process, which may not be
readily available from a TDSE solution. Another advantage of these models is that
they are often computationally much simpler than the numerical solution of the TDSE.

The advantage of solving the TDSE, on the other hand, is that it is the most rigorous tool that theorists use to predict and to validate
experimental results~(see for example Refs.~\cite{Huismans2011,les,pad_exp1,pad_exp2,kubel,PhysRevLett.126.113201}) and to compare with the predictions
of above mentioned methods (see for example Refs. \cite{lai,cqsfa,cqsfa1,jmo1,geng,arbo,maxwell}). In this sense, numerical solutions of the
TDSE are often used as a benchmark.

In this section we give a brief introduction to the numerical method for solving the TDSE within SAE
approximation and dipole approximation (for more details see Ref. \cite{benjamin}) and give guidelines how
to extract PMD from the time-dependent wave-packet calculations.

The initial state used as a starting point in the TDSE calculations is obtained
by solving the stationary Schr\"{o}dinger equation
 for an arbitrary spherically symmetric binding potential $V(\mathbf{r}) = V(r)$ in spherical coordinates:
 \begin{equation}
H_0\psi(\ver)=E\psi(\ver),\quad H_0=-\frac{1}{2}\nabla^{2}+V(r).
\end{equation}
The solution $\psi(\ver)$ can be written as
\begin{equation}
\psi_{n\ell m}(\ver)=\frac{u_{n\ell}(r)}{r}Y_\ell^m(\Omega),\quad \Omega\equiv (\theta,\varphi),
\end{equation}
where the  $Y_{\ell}^{m}(\Omega)$ are spherical harmonics. The radial function $u_{n\ell}(r)$ is a solution of the radial Schr\"{o}dinger
equation:
\begin{equation}
H_\ell(r)u_{n\ell}(r) =  E_{n\ell}u_{n\ell}(r), \label{tdse:rad}
\end{equation}
\begin{equation}
H_\ell(r)=-\frac{1}{2}\frac{d^{2}}{dr^{2}}+V(r)+\frac{\ell(\ell+1)}{2r^{2}},\label{tdse:rad1}
\end{equation}
where $n$ is the principal quantum number and $\ell$ is the orbital quantum number. The initial wave function $\psi_{n\ell m}(\ver)$ is propagated under
the influence of an intense laser field as described by the TDSE:
\begin{equation}
i\frac{\partial\Psi(\ver,t)}{\partial t}=\left[H_0+V_I(t)\right]\Psi(\ver,t),\label{tdse}
\end{equation}
where $V_I(t)=-iA(t)\partial_{z}$ is the interaction operator in the dipole approximation and velocity gauge.

We assume that the laser field is linearly
polarized along the $z$ axis, so that the vector potential is given by $A(t)=-\int^{t}E(t')dt'$, where $E(t)$ is the electric field of the laser pulse:
\begin{eqnarray}
E(t)=E_0\sin^2\left(\frac{\omega t}{2N_c}\right)\cos(\omega t),\quad t\in[0,T_p].
\end{eqnarray}
Here $E_{0}$ is the electric field amplitude, $\omega=2\pi/T$ is the laser-field frequency and $T_p=N_cT$ is the pulse duration, with $N_c$ the number of
optical cycles. At the end of the laser-atom interaction $t=T_{p}$ we obtain the time-dependent wave function $|\Psi(T_{p})\rangle$ which
contains all relevant information about the simulated process.

The question is how do we extract this information from the final wave function $|\Psi(T_{p})\rangle$? The formally exact
PMD can be extracted from $|\Psi(T_{p})\rangle$ by projecting it
onto the continuum states of the field-free Hamiltonian $H_{0}$ having the linear
momentum $\vk=(k,\Omega_\vk)$, $\Omega_\vk\equiv(\theta_\vk,\varphi_\vk)$. We
call this method the PCS (Projection onto Continuum
States) method. In a typical photoionization experiment a photoelectron ends up in quantum states with a linear momentum $\vk$, so that corresponding continuum
states must be localized in momentum space. The continuum states that describe such a quantum state
obey the so called \textit{incoming} boundary condition 
and can be written as the partial-wave expansion~\cite{roman,starace}:
\begin{equation}
\psi_{\vk}^{(-)}(\ver) = \sqrt{\frac{2}{\pi}}\frac{1}{k}\sum_{\ell,m} i^{\ell}e^{- i\Delta_{\ell}}\frac{u_{\ell}(k,r)}{r}
Y_{\ell}^{m}(\Omega)Y_{\ell}^{m*}(\Omega_\vk),\label{cont_st}
\end{equation}
where $\Delta_{\ell}$ is the scattering phase shift of the $\ell$th partial-wave. The continuum states (\ref{cont_st})
merge with the plane wave at the time
$t\rightarrow +\infty$: $\psi_\vk^{(-)}(\ver,t)\rightarrow (2\pi)^{-3/2}e^{i(\vk\cdot\ver - E_\vk t)}$.
 The probability $P(E_\vk,\theta_\vk)$ of
detecting the electron with kinetic
energy $E_{\vk}=k^{2}/2$ emitted in the direction $\theta_\vk$ is given by
\begin{equation}
P(E_\vk,\theta_\vk)=2\pi k \left|\langle \psi_{\vk}^{(-)}|\Psi(T_{p})\rangle\right|^{2},\label{pmd}
\end{equation}
where $\vk = (k_{x},k_{z}) = (k\sin\theta_{\vk},k\cos\theta_{\vk})$.

It is worthy to note that in some cases it can be cumbersome to obtain continuum states since
the continuum states are only known in an analytical form for the pure Coulomb potential.
If that is the case, an approximate PMD can be obtained by what we call the PPW (Projection onto Plane Waves) method. This approach
for obtaining the PMD from the time-dependent wave-packet calculations has been introduced and discussed in details in Ref.~\cite{madsen}.

After the laser field has been turned off, the wave function $|\Psi(T_{p})\rangle$ is propagated for a time $\tau$
under the influence of the field-free Hamiltonian $H_{0}$. The time interval $\tau$ has to be large enough so that even the slowest
part of the wave function $|\Psi(T_{p}+\tau)\rangle$ has reached the
asymptotic region $r>R$ where we can neglect
the atomic potential, $V(r)\approx 0$.
By excluding the bound part  of the wave function  $|\Psi(T_{p}+\tau)\rangle$
which we assume is spatially localized in region $r<R$, we can obtain PMD by projecting the continuum part of the wave function
onto a plane-wave:
\begin{equation}
P(E_\vk,\theta_\vk) \approx P^{\prime}(E_\vk,\theta_\vk)= 2\pi k \left|\langle \Phi_{\vk}|\Psi^{\prime}(T_{p}+\tau)\rangle\right|^{2},\label{pmd_pw}
\end{equation}
where we use plane wave $\Phi_{\vk}(\ver)$ given as the partial-wave expansion:
\begin{equation}
 \Phi_{\vk}(\ver) =\sqrt{\frac{2}{\pi}} \sum_{\ell,m}i^{\ell}j_{\ell}(kr)Y_{\ell}^{m}(\Omega)Y_{\ell}^{m*}(\Omega_\vk).
 \label{pw_exp}
\end{equation}
where $j_{\ell}(kr)$ is the spherical Bessel function of order $\ell$.
The prime on the time-dependent wave function in (\ref{pmd_pw}) indicates that we take only
part of the wave function $|\Psi(T_{p}+\tau)\rangle$ that has reached
beyond the border of the asymptotic region, $r>R$. In all presented calculations we have set $R=40$~a.u.

 \begin{figure}[t!]
 \centering
 \resizebox{0.45\textwidth}{!}{%
  \includegraphics{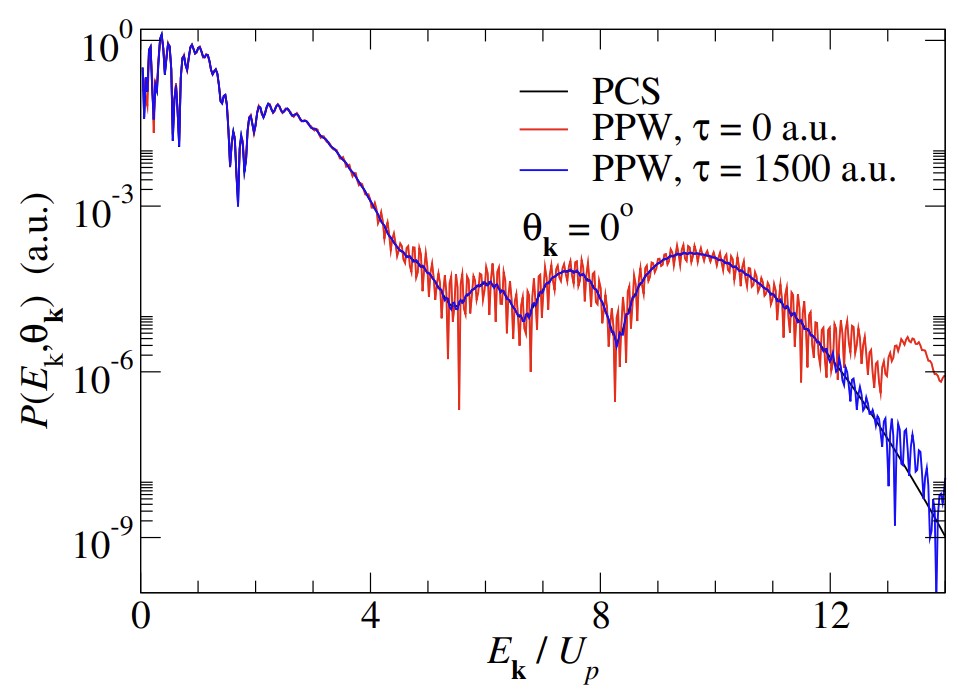}
}
  \caption{Photoelectron spectra for fluorine negative ion $\mathrm{F}^{-}$ obtained
  by the PCS and PPW methods for $\theta_{\vk}=0^{\circ}$. The exact PCS spectrum is
  depicted with the solid black line,
  while results obtained by the PPW method are depicted by the red and blue solid lines obtained
  with two different post-pulse propagation times as indicted in the legend. The laser field parameters are
  $I=1.3\times 10^{13}$~Wcm$^{-2}$,  $\lambda =1800$~nm, and $N_{c}=4$.}
  \label{fig:ppw1}
 \end{figure}

Let us now compare these two approaches for obtaining PMD. As an example we use the fluorine
negative ion $\mathrm{F}^{-}$. This choice is motivated by the fact that, after electron emission,
there is no long-range (i.e., Coulombic) potential, and, hence, the applicability of the
PPW method is clear. Within the SAE approximation we model the corresponding potential by the Green-Sellin-Zachor potential
with a polarization correction included~\cite{GSZpot}:
\begin{equation}
V(r) = -\frac{Z}{r\left[1+H\left(e^{r/D}-1\right)\right]}-\frac{\alpha}{2\left(r^2 + r_p^2\right)^{3/2}},
\end{equation}
with $Z=9$, $D=0.6708$, $H=1.6011$, $\alpha=2.002$, and $r_{p}=1.5906$. The $2p$ ground state of $\mathrm{F}^{-}$ has the electron affinity
equal to $I_{p}=3.404~$eV.  

In Fig. \ref{fig:ppw1} we show ionization probability in the direction $\theta_{\vk}=0^{\circ}$ for the
laser field intensity $I=1.3\times 10^{13}$~Wcm$^{-2}$, wavelength $\lambda =1800$~nm, and
four optical cycles laser pulse duration, $N_{c}=4$. The solid black line represent the
ionization probability obtained by the PCS method. The red line represent the ionization probability obtained by the PPW method
with the post-pulse propagation of the wave function equal to $\tau = 0$~a.u. We can see
that the low-energy part of the photoelectron spectrum agrees
quite well with the
exact result. On other hand, the high-energy part of the spectrum exhibits oscillations which are absent in the PCS results. These oscillations can
be smoothed by increasing the time of
the post-pulse propagation up to $\tau=1500$~a.u. The results
for $\tau=1500$~a.u. are depicted by the blue solid line. Our experience tell us that general rule of thumb
 is that as we increase the post-pulse propagation time,
the agreement between these two methods becomes better in the high-energy region, although this implies that we have to use a larger
spatial grids on which the TDSE is numerically solved. Therefore, one has to
compromise between the consumption of computing resources and obtaining fully converging photoelectron spectrum.

  \begin{figure}[t!]
 \centering
 \resizebox{0.5\textwidth}{!}{%
  \includegraphics{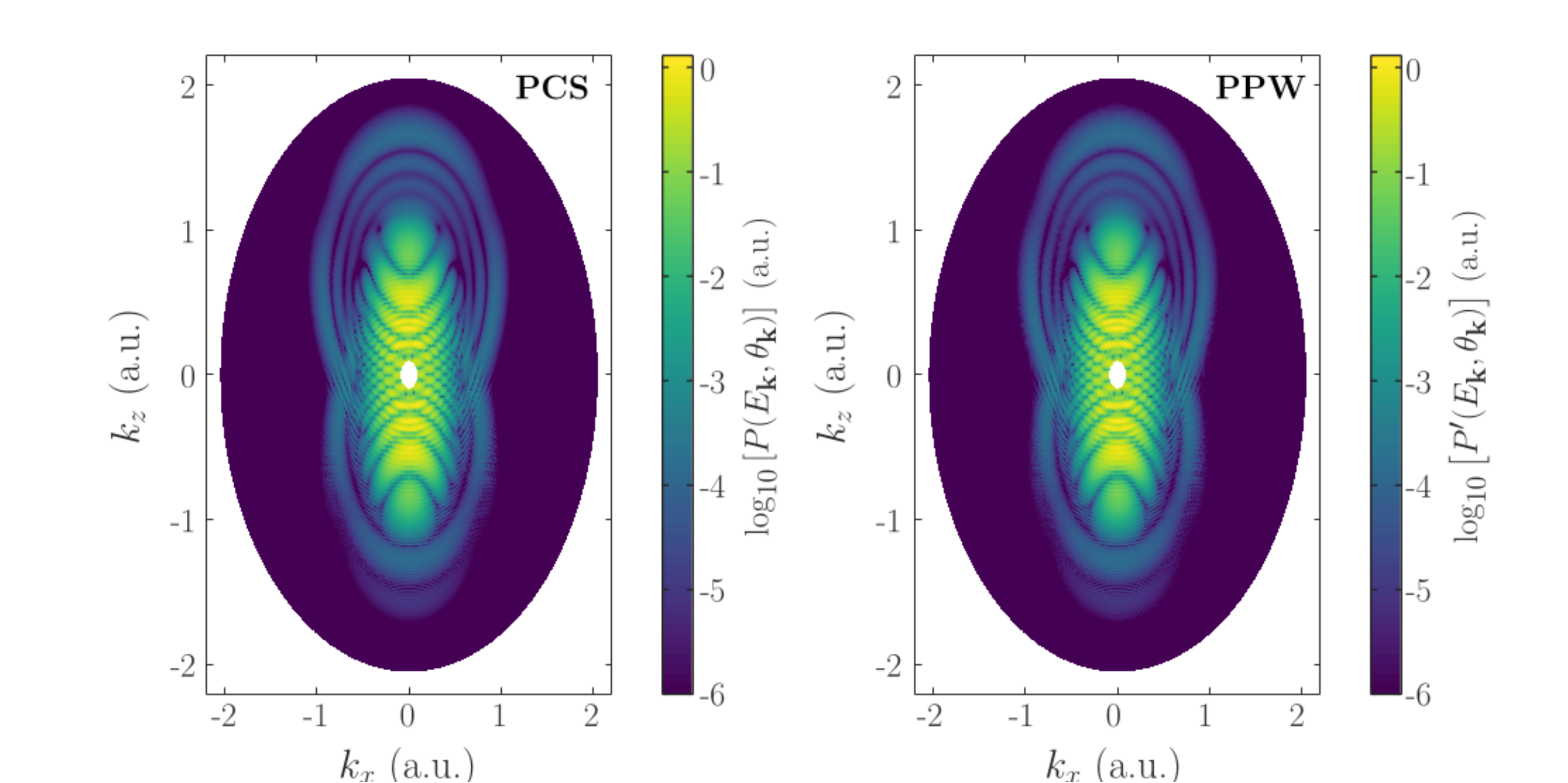}
}
  \caption{The full PMD for fluorine negative ion $\mathrm{F}^{-}$ obtained by the
  PCS and PPW methods. The left panel shows PMD obtained by the PCS method and the right
  panel shows PMD obtained by the PPW with $\tau=1500$~a.u.
 Laser field
parameters are the same as in Fig. \ref{fig:ppw1}.  }
  \label{fig:ppw2}
 \end{figure}

In Fig. \ref{fig:ppw2} we show full PMD obtained by the PCS (left panel) and PPW  methods with $\tau=1500$~a.u (right panel). The laser field
parameters are the same as in Fig. \ref{fig:ppw1}. As we can see these two methods produce identical PMDs.

In the case of atomic photoionization where the liberated electron moves in the modified
Coulomb potential, it would be the natural choice to use
the Coulomb waves as a good approximation to the true continuum states, but our calculations show that even
in the case of atomic photoionization,
the plane waves must be used as the final states of the detected electron since only the plane waves
are eigenstates of the momentum operator.
Analogous studies for atomic ionization are currently being prepared and will be published elsewhere.

\section{Coherence, decoherence and incoherence}
\label{sec:decoherence}
Coherence is defined as the capability of waves to interfere. The extension to
waves of different frequencies is the foundation of mode locking in femtosecond lasers. Interestingly, we encounter coherence in both ATI and HHG, which
are characterized by frequency combs of photoelectron or photon energies, respectively.
The time-domain flip side of the frequency comb are the “pulse trains” of the
continuum EWPs created around the peaks of the laser electric field. The sub-cycle EWPs
themselves can be understood as multi-mode interference of coherent electron waves originating
from the same laser half-cycle. In the present case of continuum wave packets, the frequency (momentum) spectrum of the wave packets is wide and continuous.

Ultrafast science is particularly interested in tracking the evolution of bound wave packets, as they allow microscopic
insights into the dynamics occurring inside atoms, molecules or solids. Typically, bound wave packets have a discrete
spectrum, implying a periodicity in time. Strong-field ionization has been shown to allow the preparation of coherences between different electronic states \cite{Rohringer2009}.  
 Well-known examples of bound electron wave packets, that have been tracked with
ultrahigh time resolution, include spin-orbit wave packets in rare-gas ions~\cite{kubel2019spatiotemporal,Fleischer2011,Goulielmakis2010} and
charge migration in polyatomic molecules~\cite{Calegari2014,Kraus2015}. Furthermore, periodic
vibrational~\cite{Ergler2006} and rotational wave packets~\cite{RoscaPruna2001,Stapelfeldt2003,Karamatskos2019} in molecules have been tracked.

Contrary to the periodic example considered above, processes such as chemical reactions are
typically non-periodic. In this context, we can consider a process as coherent if its
time evolution clearly depends on some initiating event, usually the interaction with a pump laser pulse. This is the
prerequisite to study ultrafast dynamics, for example using a pump-probe scheme.

Some processes can lead to the apparent loss of coherence, often referred to as decoherence. This may occur, for
example, in the vicinity of conical intersections, where the electron and
nuclear degrees of freedom are strongly coupled. Hence, energy stored in electronic degrees of freedom can
be transferred to vibrational motion, and the relationship between the pump event and the ensuing dynamics is lost.
A particularly interesting example is the charge migration process studied by Callegari, et al.~\cite{Calegari2014}. They
observed an oscillating ion yield due to electronic coherence immediately following XUV photoionization of phenylalanine~\cite{Calegari2014}. After a
few 10s of femtoseconds, the delay-dependent oscillations disappear and the measured signal became static. This is a clear signature of decoherence.

Generally, coherence is lost if a coherent system is coupled to a ``bath", and it is
interesting to scale the size of the bath. For example, consider electron emission from a
pair of identical atoms at a fixed internuclear distance $R$. Emission could originate from
either one of two atoms, leading to an interference term similar to the one in LIED, $\exp{\left(i k R\right)}$. In this
example, the nuclei represent the bath that may be coupled to the electron motion. Kunitski, et al., realized such an
experiment using Neon dimers exposed to an intense circularly-polarized laser field~\cite{MaksimNatureCom}. They
found that sought-after interference can be observed when they keep track of the bath, i.e. measure the nuclei in
coincidence with the electrons. Specifically, an interference structure only appears if one selects for the parity
of the ionic state on which Ne$_2^+$ dissociates~\cite{MaksimNatureCom}. Experiments on the dissociative
multiphoton ionization of H$_2$~\cite{Wu2013} can be interpreted in a similar manner. In the photoelectron spectrum, no clear
ATI peaks are seen. However, when considering the energy transferred to the nuclei, the coherent peak structure of ATI is restored. Again, the energy
absorbed by the nuclei only seemingly leads to decoherence but coherence is maintained if the measurement includes all relevant observables in the bath.

With increasing complexity, it becomes increasingly difficult to keep full track of the bath. For example, in
polyatomic molecules a plethora of nuclear degrees of freedom exists, such that it is extremely challenging to
measure all of them once nuclear motion sets in, which happens a few femtoseconds after the pump pulse. This challenge
increases even more, if one allows interaction with the environment. An illustrative example is the work of Hackerm\"uller, et al., who studied the
decoherence of matter waves due to thermal emission of radiation~\cite{hackermuller2004decoherence}: as the temperature is increased, decoherence becomes stronger due to thermal emission of radiation. Similarly, lasing or superfluorescence are impeded by spontaneous emission or nonradiative transitions \cite{Benediktovitch2019}, while decoherence of molecular rotations occurs through collisions with the environment~\cite{Milner2014,Stickler2018}. These decoherence phenomena relate to ongoing efforts to test the limits of quantum mechanics by studying interference phenomena in mesoscopic systems~\cite{Mohanty1997}.

Let us return to the prototypical examples of ATI and HHG of gaseous atoms. Despite the absence of
internal degrees of freedom that could lead to decoherence, ATI of different atoms is incoherent, i.e.~all interference
phenomena in ATI take place on the single-atom level. This is entirely different to the case of HHG where all atoms in
the focal volume radiate harmonics coherently. As a consequence, phase matching is an important issue for the case
of HHG~\cite{lewensteinmodel}, but irrelevant for ATI. Moreover, this ``macroscopic coherence" of the HHG
process is the reason for the sharp HHG peaks, while the contrast between ATI peaks is much lower. Given
their close relationship, this difference between HHG and ATI is remarkable. But what is the underlying reason? An important
difference between the two processes is the fact that ATI leads to the production of an ion, while the atom has
returned to its ground state after HHG. In other words, it is possible to tell which atom has undergone ATI,
but not which one has undergone HHG. This information is equivalent to measuring through which slit the particle
passes in Young's double slit experiment. This picture agrees well with the Neon dimer experiment discussed above: if we
know the ionic state, the interference is restored.

If the created ion destroys the coherence of ATI from multiple sources, this raises the question whether we
can make ATI coherent by studying solid state systems, where no ion is created. Suitable candidates would be
extended systems in the condensed phase, perhaps systems of nanostructures~\cite{Heimerl2020}.

\section{Toward nonlinear ultrafast spectroscopy of quantum materials}
\label{sec:material}

Recently, the HHG in solids has been attracting the attention of condensed matter
physics, see \cite{Ghimire2019} for a recent review. The effect can be observed at quite moderate
intensities below the ionization threshold. As such it allows probing solid state sample without
inflicting optical damage. HHG may be useful for the study of several transport charge and spin properties, not
only in semiconductors but also in materials which exhibit unique and novel topological effects such as 2D and 3D
topological Insulators.\cite{GhimireNatPhy2011,Liu2017,VampaPRL2014,VampaNat2015}.

The character of physical laws on the atomic scale (Angstrom and nanometer) is dramatically
different in solids than in gases~\cite{VampaNat2015,VampaJPB2017,EOsikaPRX2017,ShouCheng2011,LaughlinNobel1999,HaldaneNobel2016}. Specifically,
in the SFA, the energy of the ground state is assumed constant with respect to the momentum ${\bf p}$, while the
energy spectra of the continuum states at large distance from the parent ion, can be considered as
parabolic in ${\bf p}$~\cite{VampaJPB2017}. In condensed matter terms, the former corresponds to an infinite hole mass. Thus, in gases,
the ground state is localized, while the free continuum electron is
moving in a trajectory driven by the laser field. Re-combination may
take place at the ground state initial position in gases.  On the
contrary, in solids the situation is rather different due to periodicity
of the static potential. The possibility of electron-electron correlation
effects~\cite{Murakami2018,SilvaNatPhoton2018}, electron-phonon effects, Spin-Orbit
couplings~\cite{Angel2020,MarkusPRB2020,baykusheva2020strongfield} and spin orders'
effects~\cite{Takayoshi2019} offer a new and extremely attractive research field for
ultrafast spectroscopy and laser control. For instance, the hole can be driven by the laser
as well, since the energy dispersion of valence states is not zero~\cite{VampaJPB2017,EOsikaPRX2017}. This causes
novel ultrafast dynamics, which differ from those in the gas phase, and are present not only in ordinary
semiconductors but also in quantum materials~\cite{LaughlinNobel1999,HaldaneNobel2016,Borsch2020,ChaconPRB2020}.

Specifically, nonlinear spectroscopy has made it possible extracting information of the band structure in semiconductors
using a pump-probe scheme and associating the HHG spectra to the inter-band
mechanism~\cite{VampaNat2015}. Furthermore,
ultrafast metrology at THz frequencies has allowed for the observation of electron-hole
recombination and Bloch Oscillation at special
valley points in MoS$_2$ or WS$_2$~\cite{Ruber2018,Borsch2020}. Additionally, attosecond transient
absorption experiments for core valence electrons in solids are nowadays in the focus of
attention of attosecond condensed matter physics~\cite{Buades2019,Ruber2018}. The rapidly emerging
field of nonlinear spectroscopy in
quantum materials ~\cite{Buades2019,SilvaNatPhot2018,Bauer2018}, i.e. topological materials,
Weyl semimetals, etc.~\cite{Angel2020,ChaconPRB2020,SilvaNatPhot2018,Bauer2018} is
attracting the attention of several experimental and theoretical research groups around the world. Those materials
are extremely important, since their special features, i.e. topological conducting and isolating bands
protected by the fundamental symmetries, are robust against energy dissipation and material
perturbations~\cite{LaughlinNobel1999,HaldaneNobel2016}. These unique features promise
interesting application of topological insulators (TIs)~\cite{Zhang2009,Chen2009Science} in the
optimization of electronic devices; more precisely the transistors and the logical operations
defined in the electronic devices~\cite{Borsch2020,Lu19,Neupane2014}.

Attosecond science will expand its frontiers to new challenges, research fields and may open up new options to control
transport and optical features in quantum materials ~\cite{Borsch2020}. For example, HHG or other nonlinear optical
techniques may provide access to the electronic and dynamical properties of quantum materials. The topological
invariant defines whether or not a material is topological. It is directly linked to the electron wave function from
the crystalline structure, which characterizes the transversal current with respect to a longitudinal applied voltage, exhibiting
quantum anomalous Hall effects. The open question which still remain a challenge is how the topological invariants
might be associated with the HHG spectrum.
\begin{figure}
    \centering
    \resizebox{0.5\textwidth}{!}{\includegraphics{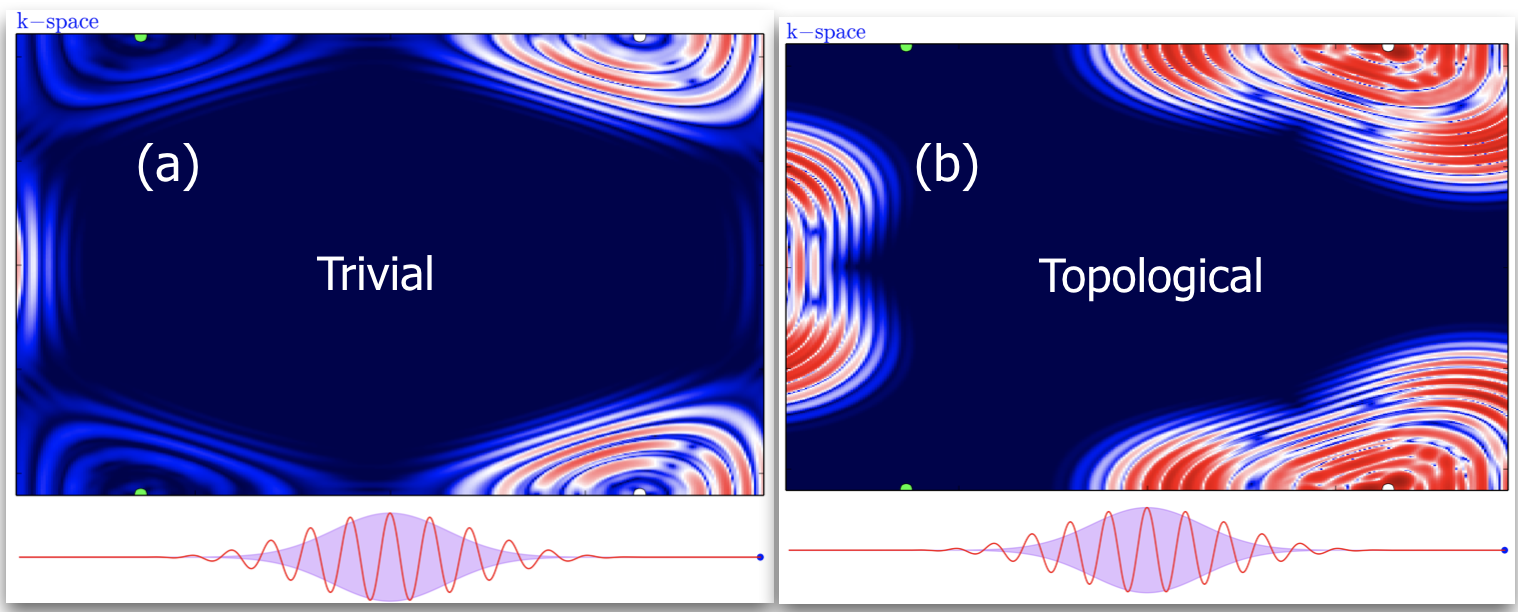}}
    \caption{Quantum interference in trivial and topological materials. Final crystal momentum distribution
    in Haldane model for the conduction band of the trivial and topological phase, are
    depicted in (a) and (b), respectively. Both, trivial and topological materials
    have the same band gap ($\varepsilon_g\approx 3.0$~eV), but exhibit totally
    different topological invariant or Chern number $\nu=0$, for the trivial
    phase, $\nu=+1$ for the topological phase. The same circularly polarized driving laser was used for both topological phases.}
    \label{fig:TOPO}
\end{figure}

We give here an example from the area of topological ultrafast nonlinear spectroscopy,
especially in the interaction of an ultrashort and intense MIR laser wavelength of 3.2~$\mu$m and
intensity of $10^{11}$~W/cm$^2$. The Haldane model (HM) is the first which appears to predict
quantum Hall effects (QHEs) without Landau levels or more precisely without magnetic fields~\cite{Haldane1988}. This model
belongs in the class of the so called Chern Insulator classification for topological materials which simply means
the Chern number, $\nu = \frac{1}{2\pi}\int_{\rm BZ} d^2{\bf k}\cdot{\bf \Omega}({\bf k})$, can be $\nu=-1,\,\,0,\,\,+1$. HM thus has
three topological different phases, i.e. for $\nu\,=\,\pm 1$ the system will show
QHEs (or transversal quantized conductivity) and, for $\nu=0$ there will not be QHEs. In Ref.~\cite{ChaconPRB2020} a full revision of
HM is done in the context of how HHG can encode the topological invariant $\nu$ by means of the circular dichroism, i.e. the produced
asymmetry photon emission yield by right- and left-circularly polarized driven lasers, also a parallel
work by Silva et al.~\cite{SilvaNatPhot2018} shown that HHG can be used to track topological phases and
transitions in the similar model, but using as an observable the helicity of the HHG produced by
linearly polarized driving lasers.

Recently, Baykusheva et al. has shown numerical results of HHG from Bi$_2$Se$_3$, a typical
3D-topological insulator (TI)~\cite{baykusheva2020strongfield}. In this work, an interesting
anomalous enhancement in the non-linear optical responses of Bi$_2$Se$_3$ was observed as  the
driving laser polarization is varied from linear to circular~\cite{ChaconPRB2020}. A theoretical method was developed, which splits the contributions from the topological surface states and bulk surface states, indicating that
the responsible mechanism of that enhancement is the spin-orbit couplings of the surfaces states.

In another example in Fig.~(\ref{fig:TOPO}), we present the final crystal momentum in the first
Brillouin zone for two different topological phases $\nu=0$ and $\nu=+1$, respectively. The green
and white points denote the K' and K points respectively. The final interferogram patterns are specially different at K-points for the topological phase with a thinner fringes for the topological phases than the trivial one.

This shows that the interesting topological features can be contained by the final momentum
distribution of the conduction bands. Note however, the question of how to measure this distribution
inside the material and how to extract the topological invariant $\nu$ are still open for ultrafast
sciences and also condensed-matter physics. Specially, and in abroad sense in quantum materials~\cite{Borsch2020} such
as Dirac and Weyl semimetals~\cite{WDSMs2018} in which two different and opposite Chern numbers
lead to the generation of Fermi arcs or pseudo-magnetic mono-poles or Weyl-fermions with
chirality features~\cite{WDSMs2018}. How nonlinear ultrafast can extract this information is
still a complete research world for both theoretical and experimental science.










\section*{Authors contributions}
All authors contributed to the preparation of the manuscript.
All authors have read and approved the final manuscript.

\section*{Acknowledgements}
The authors thank B. Cooper, A. Maxwell, and C. Faria for
organizing Quantum Battles in Attoscience, and M. Vrakking for
his valuable suggestions. K.A. acknowledges financial support from the European Research Council (788218) and the
Polish National Science Center within the project Symfonia
(2016/20/W/ST4/00314).
A.C. acknowledges Max Planck POSTECH/KOREA Research Initiative Program
[Grant No 2016K1A4A4A01922028] through the National Research Foundation of Korea (NRF)
funded by Ministry of Science, ICT \& Future Planning, partly by Korea Institute
for Advancement of Technology (KIAT) grant funded by the Korea Government (MOTIE) (P0008763),
The Competency Development Program for Industry Specialist.
S.E. and M.K. acknowledge funding by
the German Research Foundation (DFG) through priority program SPP 1840 QUTIF.
B.F. acknowledges support by the Ministry for Education, Science and Youth, Canton Sarajevo, Bosnia and Herzegovina.
M.K. acknowledges funding by the DFG under project number 437321733.

\bibliographystyle{unsrt}
\bibliography{literatur}

%
%


%
%

\end{document}